\documentclass[prb,twocolumn,showpacs,amsmath,amssymb,superscriptaddress]{revtex4}

\usepackage[dvipdfmx]{graphicx}
\usepackage{mediabb}
\usepackage{dcolumn}
\usepackage{bm}
\usepackage{color}
\usepackage{amssymb}

\begin{document}

\title{
Effective model with strong Kitaev interactions for $\alpha$-${\rm RuCl_3}$
}

\author{Takafumi Suzuki }
\author{ Sei-ichiro Suga}

\affiliation{Graduate School of Engineering, University of Hyogo, Himeji 671-2280, Japan}

\date{\today}

\begin{abstract} 
We use an exact numerical diagonalization method to calculate the dynamical spin structure factors of three {\it ab initio} models and one {\it ab initio}-guided model for a honeycomb-lattice magnet $\alpha$-RuCl$_3$. 
We also use thermal pure quantum states to calculate the temperature dependence of the heat capacity, the nearest-neighbor spin--spin correlation function, and the static spin structure factor. 
From the results obtained from these four effective models, we find that, even when the magnetic order is stabilized at low temperature, 
the intensity at the $\Gamma$ point in the dynamical spin structure factors increases with increasing nearest-neighbor spin correlation.
In addition, we find that the four models fail to explain heat-capacity measurements whereas two of the four models succeed in explaining inelastic-neutron-scattering experiments. 
In the four models, when temperature decreases, the heat capacity shows a prominent peak at a high temperature where the nearest-neighbor spin--spin correlation function increases. 
However, the peak temperature in heat capacity is too low in comparison with that observed experimentally.
To address these discrepancies, we propose an effective model that includes strong ferromagnetic Kitaev coupling, and
we show that this model quantitatively reproduces both inelastic-neutron-scattering experiments and heat-capacity measurements.
To further examine the adequacy of the proposed model, we calculate the field dependence of the polarized terahertz spectra, which reproduces the experimental results: 
the spin-gapped excitation survives up to an onset field where the magnetic order disappears and the response in the high-field region is almost linear. 
Based on these numerical results, we argue that the low-energy magnetic excitation in $\alpha$-RuCl$_3$ is mainly characterized by interactions such as off-diagonal interactions and weak Heisenberg interactions between nearest-neighbor pairs, rather than by the strong Kitaev interactions.
\end{abstract}

\pacs{71.27.+a,75.10Kt,75.70.Tj,78.70Nx}
\preprint{APS/123-QED}

\maketitle

\section{\label{sec:level1}Introduction}
The realization of spin liquids has long been one of the main challenges in condensed-matter physics.
In this context, the Kitaev model on a honeycomb lattice has attracted attention because it is exactly solvable and realizes a spin liquid in the ground state \cite{Kitaev}.
Characteristics of a Kitaev spin liquid (KSL) are manifested in the elementary excitations, which are characterized by itinerant Majorana fermions and $Z_2$ gauge fields~\cite{FengXYPRL} resulting from fractionalization of quantum spins. 
This fractionalization is confirmed by the appearance of a two-peak structure in the temperature ($T$) dependence of the heat capacity $C(T)$~\cite{Nasu1,Nasu2}.
As the temperature decreases, only the nearest-neighbor (NN) spin--spin correlation function increases, until it saturates slightly below $T=T_h$, where the higher-temperature peak appears in $C(T)$.
These features are associated with the condensation of itinerant Majorana fermions at approximately $T=T_h$.
The degrees of freedom of the $Z_2$ gauge fields freeze below $T=T_{\ell}$, where the lower-temperature peak appears in $C(T)$.
Because the entropy $(R/2) \ln 2$ is released at each peak temperature (where $R$ is the gas constant),
the two-peak structure in $C(T)$ is a hallmark of fractionalization of quantum spins into two types of Majorana fermions~\cite{Nasu1,Nasu2}.

Beyond the pure Kitaev model, numerical investigations have revealed the two-peak structure in $C(T)$ even in the magnetically ordered state
when the system is near the KSL phase \cite{Yamaji}. 
In such cases, the short- and long-range spin--spin correlation functions increase separately with decreasing temperature.
Only the short-range correlation develops at $T \approx T_h$ in $C(T)$.
The ratio $T_{\ell}/T_h$ measures the distance from the KSL phase boundary \cite{Yamaji}.

Significant efforts have been invested to obtain materials in which Kitaev physics applies.
For example, a layered honeycomb-lattice compound $\alpha$-${\rm RuCl_3}$ is considered a promising candidate for applying Kitaev physics~\cite{Plump}. 
In heat-capacity measurements for $\alpha$-${\rm RuCl_3}$, a two-peak structure appears upon decreasing the temperature~\cite{Kubota,com,Do,DHirobe}:
the broad peak implies a crossover at $T \approx  85$ K \cite{Kubota,com} or 100 K \cite{Do} and another peak derived from the magnetic ordering appears at $T_{\rm N} \approx 7$ K~\cite{Banerjee1,SYPark,Sears,Do,Banerjee2}. 
Thus the observed two-peak structure in $C(T)$ is associated with $\alpha$-RuCl$_3$ being near the KSL phase.

The emergence of Majorana fermions in $\alpha$-RuCl$_3$ has been also discussed in connection with inelastic-neutron-scattering (INS) measurements~\cite{Banerjee1,Banerjee2,Banerjee3}.
Neutron-diffraction measurements of single crystal $\alpha$-${\rm RuCl_3}$ reveal Bragg peaks related to the magnetic zigzag order at the $Y/M$ points (see Fig.~\ref{fig5}) below $T_{\rm N} \approx 7$ K~\cite{SYPark,Banerjee2}.
Below $T_{\rm N}$, a gapped spin-wave excitation reaches local minima at the $Y/M$, and $\Gamma$ points~\cite{Banerjee1,Banerjee2,Banerjee3}. 
In addition, below $T_{\rm N}$, the constant-energy cut integrated over the low-energy window results in the six-pointed star-shaped intensity centered at the $\Gamma$ point~\cite{Banerjee2,Do}. 
However, upon integrating the constant-energy cut over the high-energy window, only the intensity centered at the $\Gamma$ point appears, and this intensity survives up to $T\approx 120$ K \cite{Banerjee2}. 
The authors in Ref.~\onlinecite{Banerjee2} have pointed out that this scattering intensity around the $\Gamma$ point may originate from the continuum of Majorana excitations~\cite{Do,Knolle1,Knolle2}. 
These heat-capacity and INS experiments play key roles in verifying that $\alpha$-${\rm RuCl_3}$ is near the KSL phase.

To explain these experimental results, an effective model for $\alpha$-${\rm RuCl_3}$ is highly desirable, and
several effective models have been proposed to date ~\cite{Kim2,Kim,Banerjee1,Winter,Yadav,Banerjee2,Banerjee3,KRan,Winter2,WWang,ACatuneanu,LJanssen,Winter3}. 
Although these models explain some experimentally observed thermodynamic quantities and/or low-lying excitations, 
some of their features differ qualitatively between models.  
Thus an adequate effective model remains elusive.

To contribute to resolving this problem, we use herein an exact numerical diagonalization method to calculate the dynamical spin structure factor (DSF) and $C(T)$ using the four effective models \cite{Kim,Winter,Yadav,Winter2}
to see whether they explain the key features of the INS and $C(T)$ experiments. 
We focus on models 1--4 listed in Table I.
models 1--3 are results by {\it ab initio} calculations and model 4 is an {\it ab initio}-guided model to reproduce some aspects of the INS experiments~\cite{Banerjee2}.
We also calculate the temperature dependence of the NN spin--spin correlation function and the static spin structure factor (SSF) by using thermal pure quantum states~\cite{Hans,SS1,SS2}. 
Using the numerical results, we discuss whether the NN spin--spin correlation function and the SSF increase separately with decreasing temperature.

We find that models 1, 2, and 4 clearly reproduce the six-pointed star-shaped intensity at low energy and temperature.
The six-pointed star-shaped intensity is smeared in model 3, whereas the ground state is in the zigzag-ordered phase.
In particular, models 1-4 succeed in explaining the feature of the INS experiments concerning the large intensity around the $\Gamma$ point that survives far beyond $T_{\rm N}$.
The two-peak structure of $C(T)$ is reproduced by models 1--3, whereas only a shoulder appears in the low-temperature region for model 4.
However, in models 1--4, $T_{\rm h}$ is approximately one-third to one-twelfth of the experimentally observed value $T_{\rm h} \approx 85 $ K~\cite{Kubota,com}. 
Consequently, all four of these effective models are inadequate to explain the key features of the INS and $C(T)$ experiments.
To resolve these difficulties, we propose herein an empirical effective model for $\alpha$-RuCl$_3$ that is based on the various interactions of model 1. 
We use this empirical effective model to numerically calculate the DSF and $C(T)$ and obtain results that are consistent with experiments not only qualitatively but also quantitatively.

We further use the proposed empirical effective model to investigate the experimentally observed behavior of $\alpha$-RuCl$_3$ in a magnetic field ($H$).
In $\alpha$-RuCl$_3$, the magnetic zigzag order vanishes above $H= 8\text{--}10$ T~\cite{SHBaek,JASears,IALeahy}. 
NMR measurements have clearly shown that the peak in the relaxation rate ($T_1^{-1}$) due to the magnetic zigzag order in low fields is replaced by a spin-gap behavior ${T_1}^{-1} \propto \exp (-\Delta / T )$ for $H \gtrapprox 10$ T,
where the magnetic field is oriented perpendicular to the honeycomb plane~\cite{SHBaek}.
Moreover, the spin gap $\Delta$ increases linearly in the magnetic field.
The linear-field increase of the spin gap is also observed in electron-spin-resonance spectra~\cite{ANPonomaryov} and in polarized terahertz (THz) spectra~\cite{ZWang}. 
In the INS experiments, the magnetic order is suppressed above $H \approx 7.5$ T when the magnetic field is applied parallel to the honeycomb plane~\cite{Banerjee3}.
The most interesting point of the INS results is that, the large broad intensity centered at the $\Gamma$ point appears at $H=8$ T and $T=2$ K and the intensity profile is identical to that at $H=0$ T and $T=15$ K $> T_{\rm N}$~\cite{Banerjee3}. 
If we consider that the observed broad intensity at $H=8$ T is a signature of the excitation continuum, then the possibility of a field-induced gapped spin liquid arises.
Of these experiments, we focus on INS and polarized THz spectroscopy and calculate the DSF and the polarized THz spectra of the proposed empirical effective model with magnetic fields.

The remainder of the paper is organized as follows:
 Section II introduces the details of numerical methods.
Section III shows the static and dynamical spin structure factors and compares them with the INS measurements on $\alpha$-${\rm RuCl_3}$.
In addition, we calculate the temperature dependence of the NN spin--spin correlation function and the SSF and find that with models 1, 3, and 4, 
the NN spin correlation and the long-range spin correlation increase separately, which is a characteristic of the proximity of the KSL phase~\cite{Yamaji}.
The temperature dependence of the heat capacity is also shown to investigate whether the two-peak structure is present.
We compare these calculations with the experimental results and find that for each effective model 1--4, $T_{\rm h}$ is too low.
Section IV proposes an empirical effective model that explains both the INS and the $C(T)$ experiments, and we use this model to calculate the field dependence of the polarized THz spectra.
We find that the calculated spectra are consistent with experiments~\cite{ZWang}. 
Finally, we discuss and summarize the results in Sec. V.

\section{\label{sec:level2}Model and Method}
\begin{figure}[htb]
\begin{center}
\includegraphics[width=0.95\hsize]{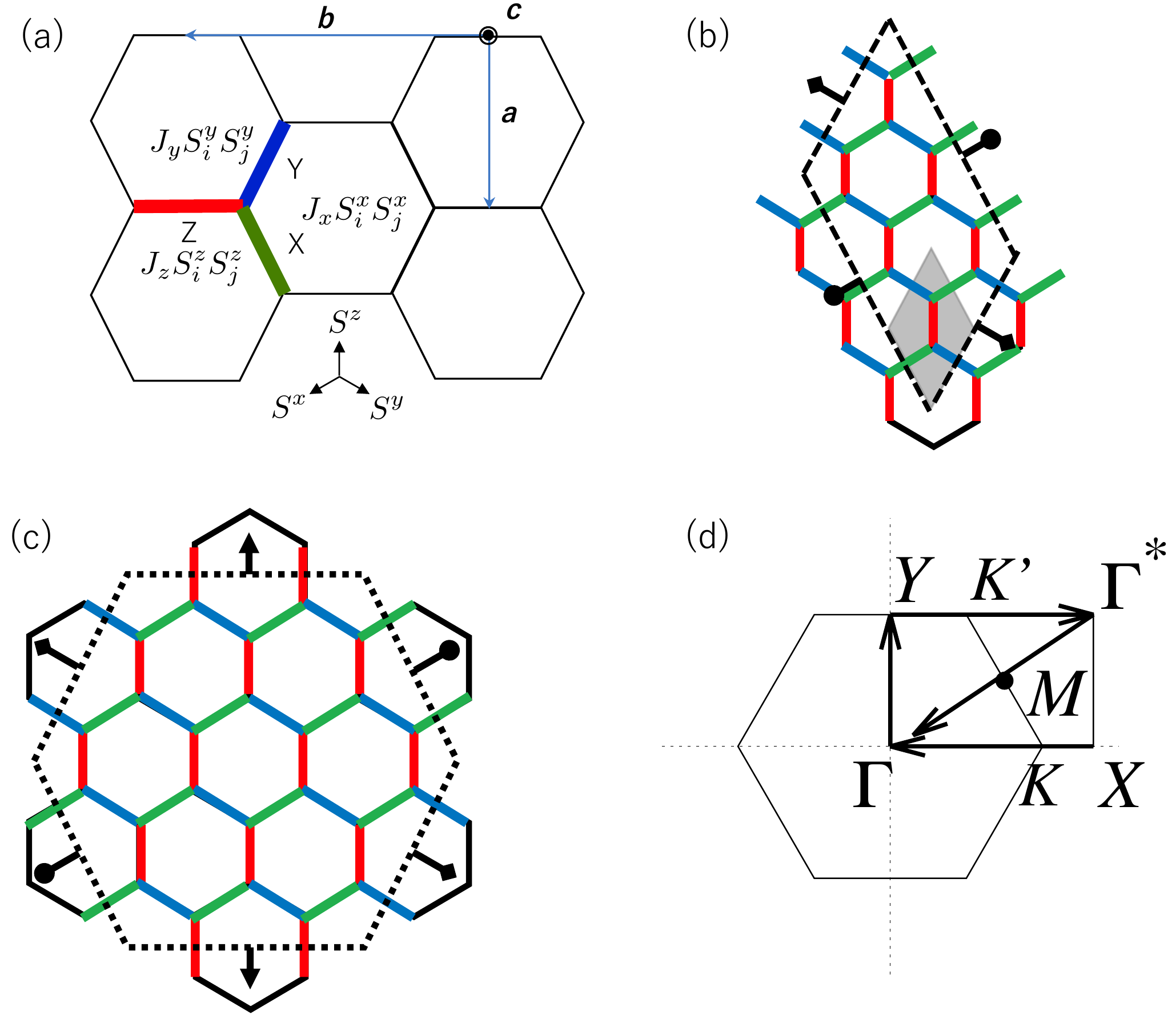}
\caption{\label{fig5} (a) Kitaev-type anisotropic interactions on a honeycomb lattice. ${\boldsymbol a}$ and ${\boldsymbol b}$ represent the reciprocal vectors for a four-sublattice magnetic unit cell. 
${\boldsymbol c}$ is oriented perpendicular to the honeycomb plane.
(b) $N=12$ site and (c) $N=24$ site clusters. 
The hatched diamond in panel (b) represents a unit cell.
Periodic boundary conditions are applied on the dotted lines with common symbols.
(d) Brillouin zone with symmetric points labeled. 
The Bragg peaks at the $\Gamma$, $Y$, $\Gamma$', and $X$ points characterize the symmetry breaking of the ferromagnetic, zigzag, N\'eel, and stripy order, respectively.}
\end{center}
\end{figure}

The Hamiltonian of the $S=1/2$ generalized Kitaev--Heisenberg model on a honeycomb lattice is given by 
\begin{eqnarray}
\label{Ham2}
{\mathcal H}_{\rm KH}=
\displaystyle\sum_p \sum_{B_p} \sum_{( ij ) \in B_p} \sum_{\mu,\nu=x,y,z} S^{\mu}_i \hat{\mathcal J}^{\mu \nu}_{B_p} S^{\nu}_j,
\end{eqnarray}
where 
$\hat{\mathcal J}^{\mu\nu}_{B_p}$ represents a $3\times3$ matrix expressing the exchange coupling between the $p$th neighboring $i$ and $j$ sites in the bond ${B_p}$.  
For instance, the matrix elements of the NN pairs on a $Z$ bond [see Fig. \ref{fig5}(a)] are 
\begin{equation}
\hat{\mathcal J}^{\mu \nu}_{Z_1}=
\begin{pmatrix}
{J_{\rm 1st}}^z &  {\Gamma_{\rm 1st}}^z  & {\Gamma'_{\rm 1st}}^z\\
{\Gamma_{\rm 1st}}^z  & {J_{\rm 1st}}^z & {\Gamma'_{\rm 1st}}^z\\
{\Gamma'_{\rm 1st}}^z  & {\Gamma'_{\rm 1st}}^z & {J_{\rm 1st}}^z + {K_{\rm 1st}}^z\\
\end{pmatrix},
\end{equation}
where $J_p^{\mu}$ and $K_p^{\mu}$ denote the coupling constants of the Heisenberg and Kitaev interactions, respectively.
The off-diagonal elements $\Gamma_p^{\mu}$ and ${\Gamma'_p}^{\mu}$ 
originate from the symmetry breaking of the crystal structure due to lattice distortions. 
In this paper, we focus on four effective models for $\alpha$-RuCl$_3$.
Three of these models (models 1--3) are solved by {\it ab initio} calculations~\cite{Kim,Winter,Yadav}, and the interactions in the fourth model are guided by {\it ab initio} calculations~\cite{Winter2}.
Table~I summarizes the details of the coupling constants for the interactions.

\begin{table*}[thb]
 \caption{\label{table1} Coupling constants for the four effective models 1--4. $J_p^{\mu}$ denotes the coupling constants of the Heisenberg-type interaction 
and $K_p^{\mu}$ denotes the coupling constants of the Kitaev-type interaction. 
$\Gamma_{p}^{\mu}$ and $\Gamma_{p}^{\prime \mu}$ are symmetric off-diagonal components of the matrix expression $\hat{\mathcal J}^{\mu \nu}_{B_p}$ in Eq. (2).
Ferromagnetic (antiferromagnetic) interactions are represented by negative (positive) values. Energy is expressed in meV.
}
\begin{tabular}{lcccccccccc}
\hline \hline 
            & $J_{\rm 1st}^{x/y}$ & $J_{\rm 1st}^{z}$ & $K_{\rm 1st}^{x/y}$ & $K_{\rm 1st}^{z}$  
            & $\Gamma_{\rm 1st}^{x/y}$ & $\Gamma_{\rm 1st}^{z}$  
            & $\Gamma_{\rm 1st}^{\prime x/y}$ & $\Gamma_{\rm 1st}^{\prime z}$  
            & $J_{\rm 2nd}$ & $J_{\rm 3rd}$ \\
            \hline
model 1~\cite{Kim} & -1.55 & -1.49 & -6.47 & -6.71 & 5.24 & 5.28 & -1.08 & -0.69 & 0 & 0 \\
model 2~\cite{Winter} & -1.7 & -1.7  & -6.7 & -6.7 & 6.6 & 6.6 & -0.9 & -0.9 & 0 & 2.7 \\ 
model 3~\cite{Yadav,THANKS}  & 1.2 & 1.2 & -5.6 & -5.6 & 1.2 & 1.2 & -0.7 & -0.7 & 0.25 & 0.25 \\
model 4~\cite{Winter2} & -0.5 & -0.5  & -5 & -5 & 2.5 & 2.5 & 0 & 0 & 0 & 0.5 \\ 
\hline\hline
\\
\end{tabular}
\end{table*}

For models 1--4, we calculate the DSF, which is defined as 
\begin{widetext}
\begin{eqnarray}
\label{sqw_ex}
S^{\mu \nu}(\mbox{\boldmath $Q$},\omega;T) \equiv -\frac{1}{Z(T)} \sum_{n,m} \lim_{\epsilon \rightarrow +0} \frac{1}{\pi} {\rm Im} \frac{\langle \phi_n |    \hat{S}^{\mu \dagger}_{\mbox{\boldmath $Q$}} | \phi_m \rangle \langle \phi_m | \hat{S}^{\nu}_{\mbox{\boldmath $Q$}}|\phi_n\rangle}{\omega+E_n+i\epsilon- E_m} e^{-(E_m-E_n)/k_BT} ,\; \mu,\nu=x,y,z,
\end{eqnarray}
\end{widetext}
where $Z(T)$ is a partition function, $\phi_n$ denotes an eigenstate of ${\mathcal H}_{\rm KH}$ with eigenvalue $E_n$, 
and $\hat{S}^{\nu}_{\mbox{\boldmath $Q$}} = N^{-1}\sum_{\mbox{\boldmath $r$}} {S}^{\nu} \exp(-i \mbox{\boldmath $Q$}\cdot \mbox{\boldmath $r$})$, with $N$ being the number of sites.  
Because we need all eigenstates and eigenvalues to calculate the DFS at finite temperature, 
the system size is limited to a small cluster. 
In this study, we calculate the DSF at finite temperatures for the cluster with $N=12$ sites shown in Fig. \ref{fig5}(b).
At $T=0$, the expression (\ref{sqw_ex})  reduces to 
$
S^{\mu \nu}(\mbox{\boldmath $Q$},\omega;T=0) \equiv -\pi^{-1} \lim_{\epsilon \rightarrow +0} {\rm Im} \langle \phi_0 | \left[ \hat{S}^{\mu \dagger}_{\mbox{\boldmath $Q$}}  \hat{S}^{\nu}_{\mbox{\boldmath $Q$}}/\left( \omega+E_0+i\epsilon-{\mathcal H}_{\rm KH} \right) \right]|\phi_0\rangle,
$
where $\phi_0$ is the ground state of ${\mathcal H}_{\rm KH}$ with energy $E_0$.  
The quantities $\phi_0$ and $E_0$ are calculated by using the Lanczos method, and then $S^{\mu \nu}(\mbox{\boldmath $Q$},\omega;T=0)$ is obtained by a continued-fraction expansion~\cite{Gagliano}. 
We calculate the DSF at $T=0$ for the cluster with $N=24$ sites shown in Fig. \ref{fig5}(c) and evaluate the sum of the diagonal elements, $S(\mbox{\boldmath $Q$},\omega;T) \equiv \sum_{\mu=x,y,z}S^{\mu \mu}(\mbox{\boldmath $Q$},\omega;T)$. 
The scattering intensity ${\mathcal I}({\boldsymbol Q},\omega;T)$ observed in the INS experiments~\cite{Do,Banerjee2} is connected with the DSFs via the form factor $f({\boldsymbol Q})^2$, namely, ${\mathcal I}({\boldsymbol Q},\omega;T) \propto f(Q)^2\sum_{\mu\nu} (1-Q_{\mu}Q_{\nu}/{\boldsymbol Q}^2)S^{\mu\nu}({\boldsymbol Q,\omega;T})$.
In the present calculation, we evaluate the form factor $f({\boldsymbol Q})^2$ for Ru$^{3+}$ from Table 2 in Ref.~\onlinecite{DTCromer}.

To discuss the thermal properties of models 1--4, we calculate the temperature dependence of the heat capacity $C(T)$, the longitudinal component $R^z(T)$ of the NN spin--spin correlation function on the $Z$ bond, 
and the SSF $S^{x/z}_{\boldsymbol Q}(T)$ by using thermal pure quantum states~\cite{SS1,SS2} for the cluster with $N=24$ sites. 
$R^z(T)$ and $S^{x/z}_{\boldsymbol Q}(T)$ are defined as
\begin{eqnarray}
R^z (T)&=&\sum_{( ij )\in Z{\rm-bond}} \langle S^z_iS^z_j\rangle \nonumber\\
&=& \frac{1}{N_b Z(T)} {\rm Tr } \left[ \sum_{( ij ) \in Z{\rm-bond}} S^z_iS^z_j e^{-{\mathcal H }/k_BT} \right]
\end{eqnarray}
and
\begin{eqnarray}
S^\mu_{\boldsymbol Q}(T)&=& \langle \hat{S}^{\mu}_{\boldsymbol Q} \rangle,
\end{eqnarray}
where $\langle \cdots \rangle$ denotes the thermal average, $N_b$ is the number of the $Z$ bond, and 
the sum in $R^z(T)$ is taken over pair sites in all $Z$ bonds.
Because the symmetries of models 1--4 guarantee $S^{x}_{\boldsymbol Q}(T)=S^{y}_{\boldsymbol Q}(T)$, 
we calculate $S^{x}_{\boldsymbol Q}(T)$ and $S^{z}_{\boldsymbol Q}(T)$.

$R^z(T)$ is a good indicator of the growth of the NN spin correlation because the strongest interaction in models 1--4 is the Kitaev coupling through the $Z$ bond.

\section{\label{sec:level3}Numerical results and discussions}
\subsection{Static and dynamical spin structure factors at $T=0$}
%
\begin{figure}[htb]
\begin{center}
\includegraphics[width=\hsize]{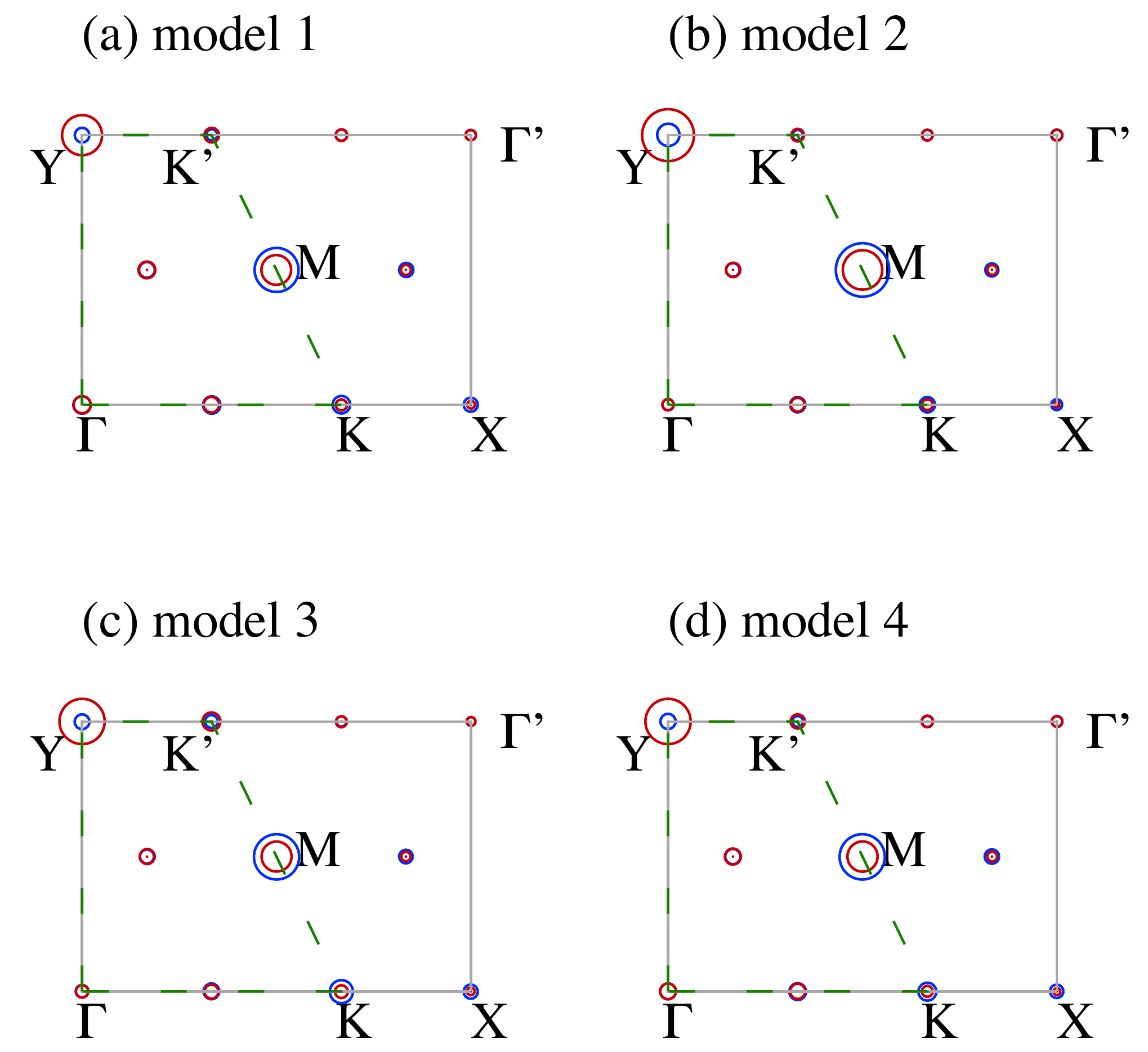}
\hspace{0pc}
\caption{\label{fig_sq} (Color online) Static spin structure factor at $T=0$ for $N=24$ cluster. Rectangular surrounded by $\Gamma\text{-}Y\text{-}\Gamma'\text{-}X\text{-}\Gamma$ denotes the Brillouin zone for a four-sublattice magnetic unit cell (see Fig. {\ref{fig5}}(d)). 
Red circles denote the transverse component $S^x_{\boldsymbol Q}(T=0)$, whereas blue circles are the longitudinal component $S^z_{\boldsymbol Q}(T=0)$.
The area of each circle is proportional to the intensity of $S^{x/z}_{\boldsymbol Q}(T=0)$ at the wave vector ${\boldsymbol Q}$.} 
\end{center}
\end{figure}

\begin{figure}[htb]
\begin{center}
\includegraphics[width=0.95\hsize]{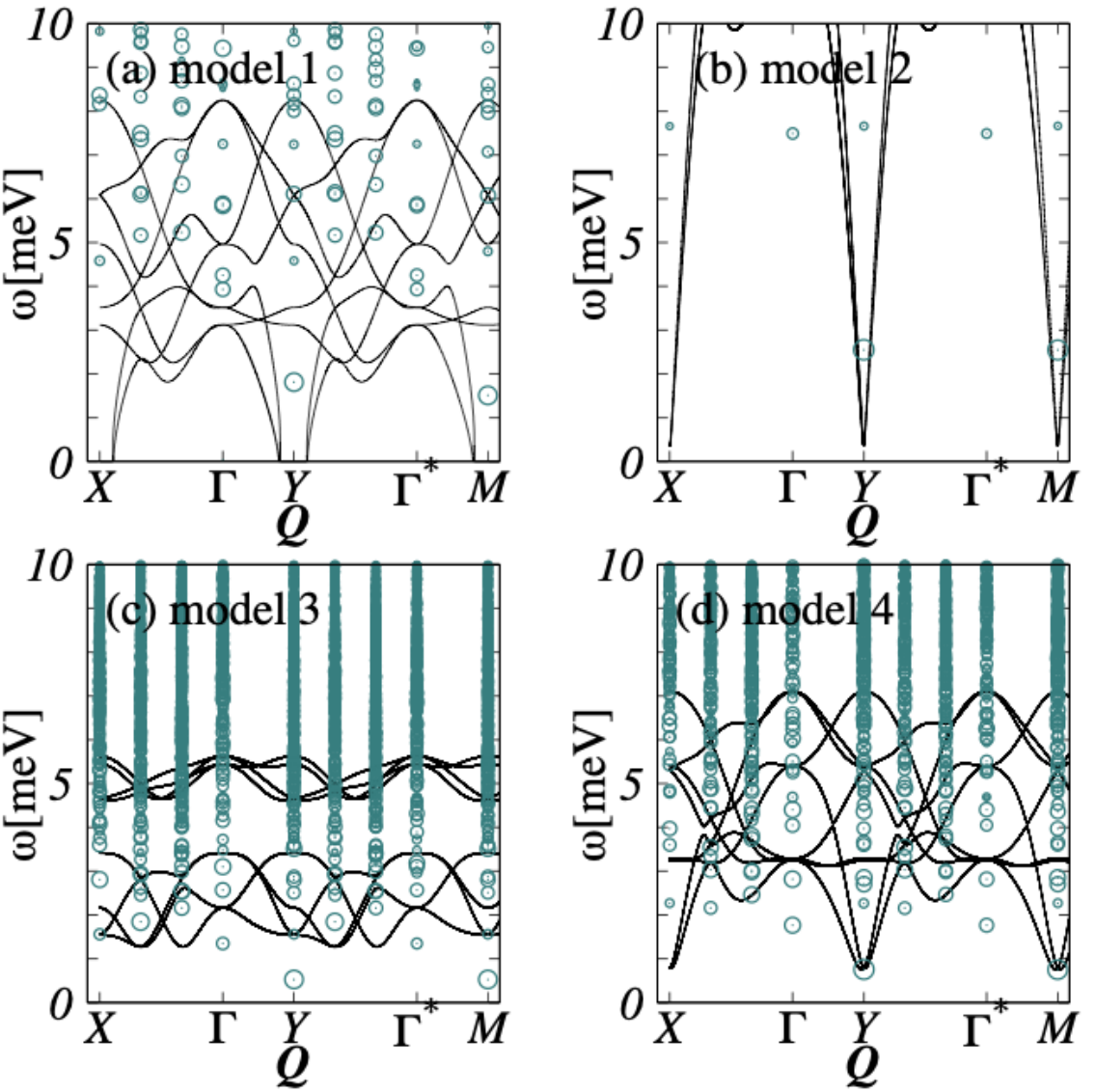}
\hspace{0pc}
\vspace{0pc}
\caption{\label{fig1} DSF $S({\boldsymbol Q},\omega;T=0)$ for $N=24$ cluster.  
The area of the circle represents $\log S({\boldsymbol Q},\omega;T=0)$.  
The solid curves are dispersion curves calculated from linearized spin-wave theory. 
For the calculations based on linearized spin-wave theory, we start from  the lowest energy state in the collinear zigzag configurations. 
The horizontal axes in panels (a)--(d) run along the arrows shown in Fig.~\ref{fig5}(d).
We set the half width of the Lorentzian to 0.001 meV.}
\end {center}
\end{figure}
%
\begin{figure}[htb]
\begin{center}
\includegraphics[width=0.95\hsize]{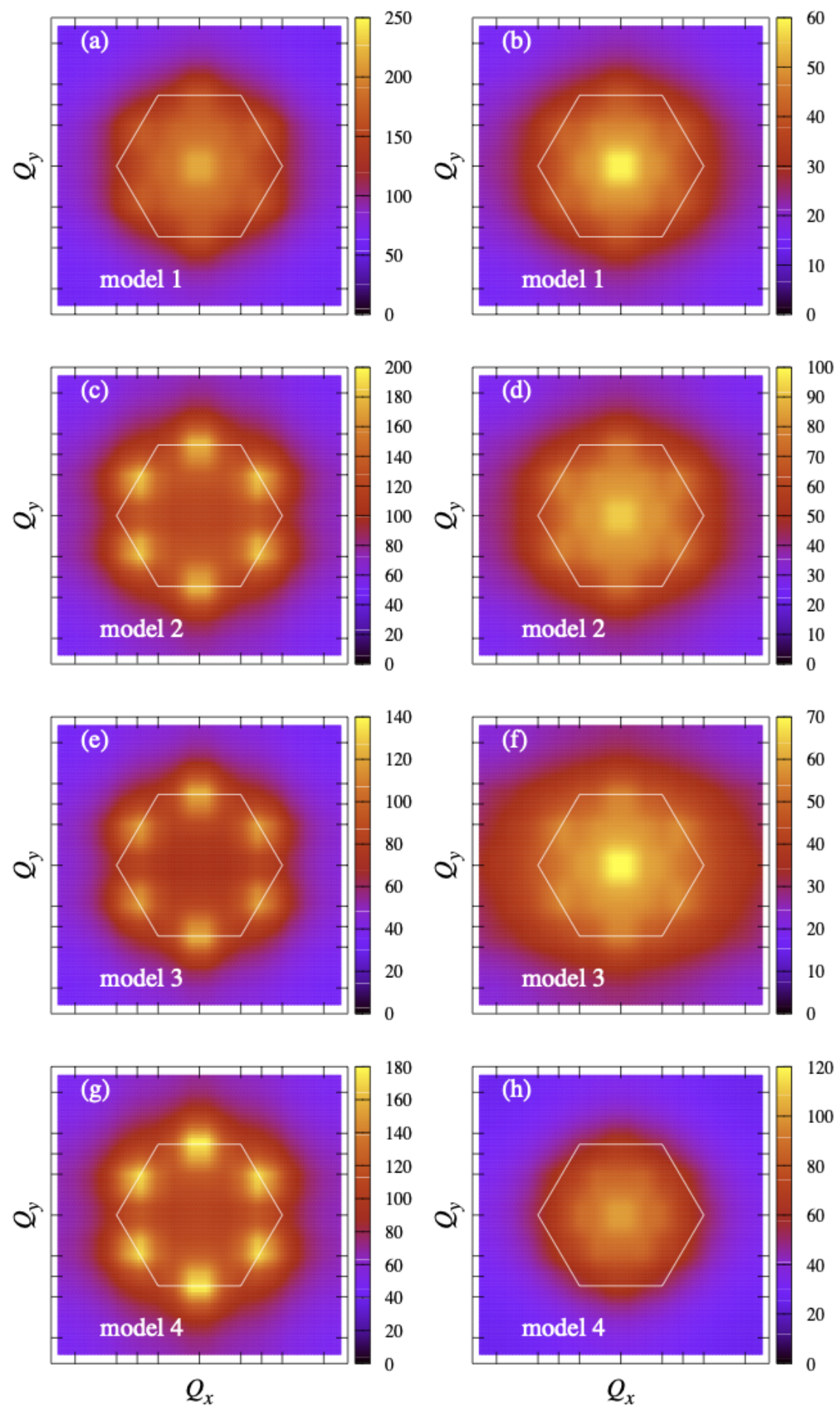}
\hspace{0pc}
\caption{\label{fig4} Constant-energy cuts of DSFs for $N=24$ cluster. 
We set the half width of the Lorentzian to 0.25 meV.
The left (right) panels show the results for integration of the DSFs for models 1--4 over the energy window [1.5,3] meV ([9,12] meV).
The momentum dependence is obtained by interpolating $S({\boldsymbol Q},\omega;T=0)$ over discretized momenta.}
\end{center}
\end{figure}

To confirm the ground states of models 1--4, we show the SSFs $S^{x}_{\boldsymbol Q}(T=0)$ and $S^{z}_{\boldsymbol Q}(T=0)$ in Fig.~\ref{fig_sq}.
In the four sub-lattice magnetic order, the peaks at the $\Gamma$, $Y$, $\Gamma'$, and $X$ points in the SSFs
characterize the ferromagnetic, zigzag, N\'eel and stripy order, respectively.
In models 1--4, the largest intensity appears at the $Y$ or $M$ point.
The intensity of $S^{x}_{\boldsymbol Q}(T=0)$ at the $Y$ point is almost identical to or slightly larger than that of $S^{z}_{\boldsymbol Q}(T=0)$ at the $M$ point. 
Therefore, the ground states of models 1--4 are in the zigzag-ordered phase, in agreement with experiment.

In Fig. \ref{fig1}, we show the DSFs at $T=0$, where the intensity is expressed by the area of the circle on a logarithmic scale, $\log S({\boldsymbol Q},\omega;T=0)$. 
The solid curves are dispersion curves calculated based on linearized spin-wave theory (LSWT). 
In models 1--4, the largest intensities appear at the $Y/M$ points and at $\omega \approx$ 1.8,  2.6, 0.5, and 0.8 meV, respectively.
These large intensities stem from the magnetic zigzag order in agreement with the experiments~\cite{Do,Banerjee1,Banerjee2}.

The LSWT results for model 1 fail to explain the DSF. 
To calculate the spin-wave dispersion, we assume the lowest energy state in the collinear zigzag configurations. 
This assumption yields a negative energy, indicating that the lowest-energy state for the given parameter set is not in the zigzag phase.
The use of model 2 makes it difficult to conclude whether the LSWT captures the low-lying excitation in the DSF.
Most poles in the DSF exist for $\omega \gtrapprox 12$ meV, which is caused by the large third-neighbor Heisenberg interaction $J_3$.  
The LSWT results for models 3 and 4 seem to explain the low-energy excitation of the DSF.
 However, there are small discrepancies between the low-lying excitations in the DSF and LSWT results.

Based on INS experiments, the authors in Refs. \onlinecite{Do} and \onlinecite{Banerjee2} have reported the characteristic features in the constant-energy cuts of the scattering intensity profile.
Figure \ref{fig4} shows the constant-energy cuts integrated over the low-energy window [1.5, 3] meV and the high-energy window [9,12] meV of the scattering intensity ${\mathcal I}({\boldsymbol Q},\omega;T=0)$ for the $N=24$ cluster.
These windows are comparable to those used in the INS experiments~\cite{Do,Banerjee2}.
Note that we apply a Lorentzian with a wider half width (0.25 meV) than that used in Fig.~\ref{fig1}
because this half width is comparable to the resolution in the INS experiments~\cite{Do,Banerjee2}.

First, we focus on the constant-energy cuts integrated over [1.5, 3] meV.
The results appear in Figs. \ref{fig4}(a), \ref{fig4}(c), \ref{fig4}(e), and \ref{fig4}(g).
In models 1, 2, and 4, the dominant intensity appears at the $\Gamma$ point and the satellite peaks appear at the $Y$ and $M$ points.
Thus the six-pointed star-shaped profile is well reproduced by models 1, 2, and 4.
In contrast, the six-pointed star shape in model 3 is smeared in spite of the zigzag magnetic order. 
The lack of the six-pointed star shape in model 3 results from the range of the energy window and multiplying the form factor. 
In the present resolution, model 3 fails in reproducing the six-pointed star-shape profile.
Therefore, models 1, 2, and 4 explain the characteristics of the constant-energy cuts integrated over the low-energy window in the INS experiments~\cite{Do,Banerjee2}, whereas model 3 does not.

Next, we focus on the constant-energy cuts in the high-energy window.
Figures \ref{fig4}(b), \ref{fig4}(d), \ref{fig4}(f), and \ref{fig4}(h) show the constant-energy cuts integrated over [9,12] meV.
In the experiments, the dominant intensity of the constant-energy cuts integrated over the high-energy window is distributed around the $\Gamma$ point.
The six-pointed star-shaped profile of the intensity disappears~\cite{Do,Banerjee2}. 
For model 2, the six-pointed star-shaped profile survives, so this result is inconsistent with the INS experiments.
In contrast, for models 1, 3 and 4, the intensity at the $Y$ and $M$ point is strongly suppressed in comparison with the result for model 2,
although the tails of the intensity from the $\Gamma$ point remain near the $Y$ and $M$ points.
Based on these results, we conclude that the constant-energy cuts for models 1 and 4 partly reproduce the INS results.
Note that the intensity profile is qualitatively unchanged upon changing the energy window for integrating $S({\boldsymbol Q},\omega;T=0)$ in the range of about 10\% around $\omega \approx 10$ meV.

\subsection{Temperature dependence of dynamical spin structure factor at $\Gamma$ point}
%
\begin{figure}[htb]
\begin{center}
\includegraphics[width=0.9\hsize]{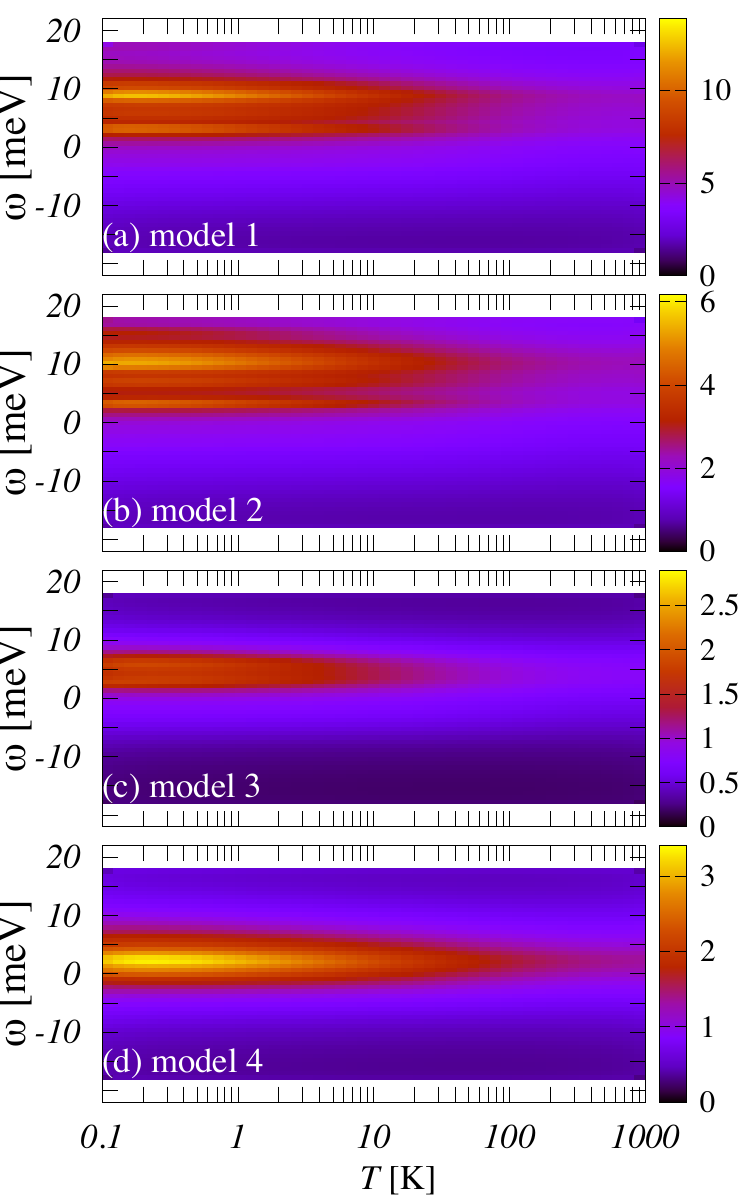}
\hspace{0pc}
\caption{\label{fig2}$S(\mbox{\boldmath $0$},\omega;T)$ at finite temperatures for $N=12$ cluster. We set the half width of the Gaussian peak to 0.25 meV.}
\end{center}
\end{figure}

In the INS experiments, the constant-energy cuts integrated over the high-energy window show that the $\Gamma$-point intensity survives up to 100--120 K~\cite{Do,Banerjee2}.
To see the thermal stability of the $\Gamma$-point intensity, we show the temperature dependence of the DSFs at the $\Gamma$ point in Fig. \ref{fig2}. 
In the finite-temperature calculations, we use the $N=12$ site cluster with low geometric symmetry shown in Fig. \ref{fig5}(b) and a larger half width (0.25 meV) for each Gaussian peak. 
Thus the resolution with respect to $\omega$ is lower compared with that for $S(\mbox{\boldmath $0$},\omega;0)$ shown in Fig. \ref{fig1}. 
Actually, the peaks at $T=0$ are smeared at finite temperatures in models 3 and 4, whereas several peaks appear at finite temperatures in models 1 and 2.

Figure \ref{fig2} shows that the $\Gamma$-point intensity in each model increases with decreasing temperature, at least below $T \approx 50$ K.
This temperature scale is of the same order as the energy scale for the largest interaction $K_{\rm 1st}^{z}\approx$ 78, 78, 65, and 58 K for models 1--4, respectively.
At approximately these temperatures, the longitudinal component $R^z(T)$ of the NN spin--spin correlation function begins to grow, as shown in Fig. \ref{fig3}. 
We thus consider that the growth of the $\Gamma$-point intensity is associated with the growth of the NN spin correlation, which is controlled by the largest Kitaev interaction $K_{\rm 1st}^{z}$. 

In the pure Kitaev model, the $\Gamma$-point intensity increases from $\omega=0$ to the nonzero $\omega$ approximately at the temperature where the NN spin correlation begins to increase with decreasing temperature \cite{JYoshitake1,JYoshitake2}. 
We consider that the same situation also happens in the present system, although the growth of the $\Gamma$-point intensity with respect to $\omega$ seems to be absent owing to the sparse energy levels coming from the small-sized cluster.

\subsection{Thermal properties}
\begin{figure}[htb]
\begin{center}
\includegraphics[width=0.75\hsize]{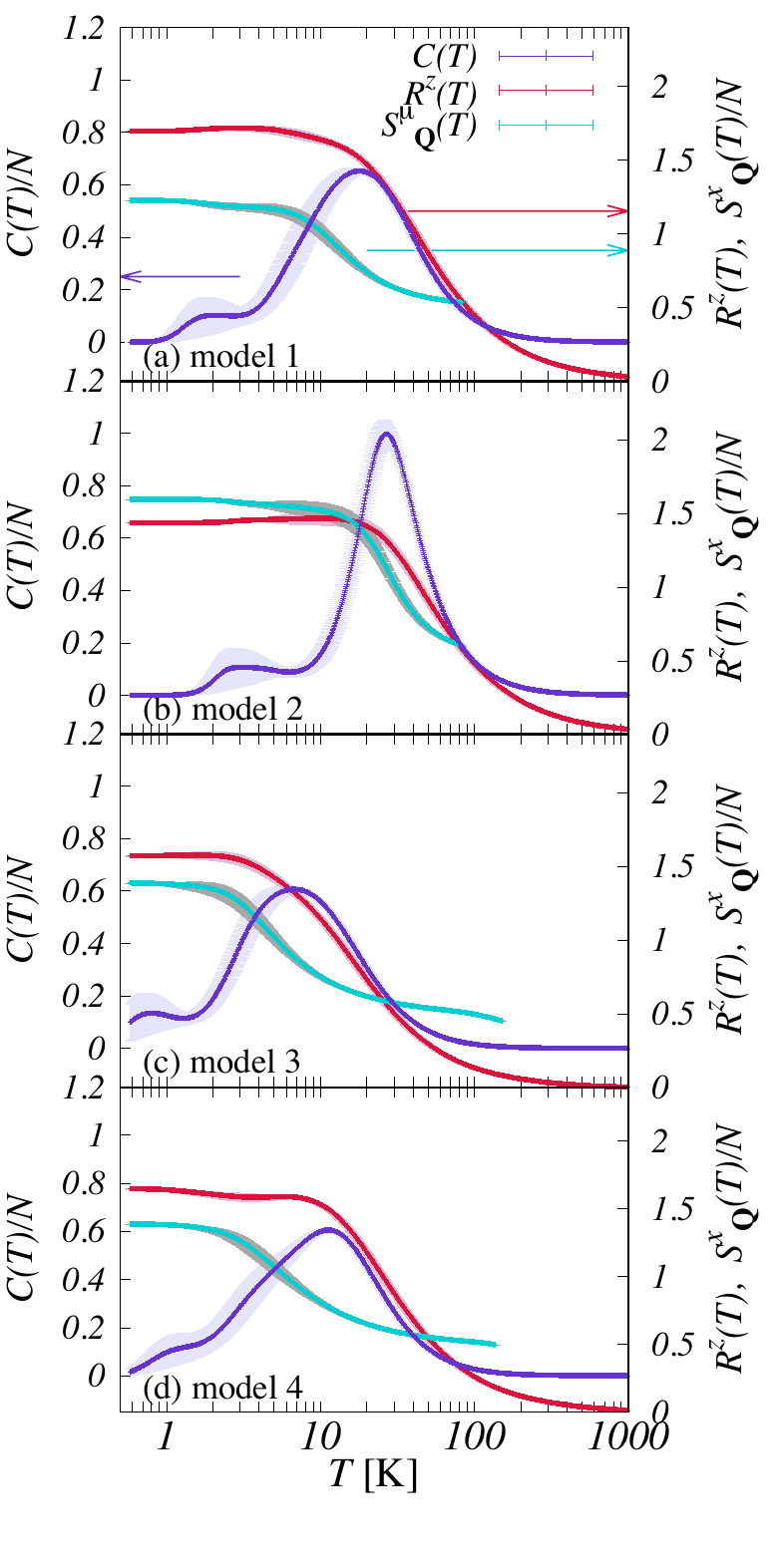}
\hspace{0pc}
\caption{\label{fig3} Temperature dependence of heat capacity $C(T)$, longitudinal component $R^z(T)$ of NN spin--spin correlation function, 
and static spin structure factor $S^{x}_{\boldsymbol Q}(T)$ at the Y point for the $N=24$ site cluster by using thermal pure quantum states~\cite{SS1,SS2}.
All results are drawn with error bars given by the variance estimate by the typical pure state~\cite{SS2,YSK}. }
\end{center}
\end{figure}

Figure~\ref{fig3} shows the temperature dependences of the heat capacity $C(T)$, the longitudinal component $R^z(T)$ of the NN spin-spin correlation function, 
and the SSF $S^{x}_{\boldsymbol Q}(T)$ at the magnetic ordering wave vector for the $N=24$ cluster. 
We show $S^x_{\boldsymbol Q}(T)$ at the $Y$ point for models 1--4 because their intensities are largest at $T=0$.

$R^z(T)$ and $S^{x}_{\boldsymbol Q}(T)$ in models 1 and 4 increase at the different temperatures, 
which means that the short- and long-range spin correlation increase separately with decreasing temperature.
We consider that model 3 still shows the separate growth of the short- and long-range spin correlation, 
although the separation is not clear in comparison to that in models 1 and 4.
The separate growth of these spin--spin correlation functions occurs in the pure Kitaev model~\cite{Nasu2} and in the Kitaev--Heisenberg model near the KSL phase~\cite{Yamaji}. 
Thus the separate growth observed here is a feature of the frustrated system that originates from the strong Kitaev interaction.  
Conversely, $R^z(T)$ and $S^{x}_{\boldsymbol Q}(T)$ at the $Y$ point in model 2 increase at approximately the same temperature. 
Because the strong third-neighbor Heisenberg interaction stabilizes the magnetic zigzag order, 
the system distances itself from the phase boundary with the KSL phase.
When the Kitaev--Heisenberg model is deeply in the magnetically ordered phase, 
the NN and long-range spin--spin correlation functions develop simultaneously with decreasing temperature~\cite{Yamaji}.

For models 1--3, $C(T)$ exhibits the two-peak structure. 
In contrast, for model 4, only a shoulder appears in the low-temperature region. 
This is because of the partially covering of the low-temperature peak by the tail of the broad higher-temperature peak. 
Following the criterion that was elucidated in Ref. \onlinecite{Yamaji}, we evaluate the ratio of the temperature of the low-temperature peak to that of the high-temperature peak: ($T_{\rm \ell}/T_{\rm h}$). 
We obtain $T_{\rm \ell}/T_{\rm h}\approx$ 0.08, 0.1, and 0.11 from models 1--3, respectively, which means that model 1 is nearest the KSL phase, followed by model 2 and then model 3.
The ratios are approximately the same as that observed for $\alpha$-${\rm RuCl_3}$: $T_{\rm \ell}/T_{\rm h} \approx 0.09$ \cite{Kubota,com} or 0.065 \cite{Do}.  
However, the values of $T_{\rm h}$ for models 1--4 are $T_{\rm h} \approx$ 20, 28, 7, and 12 K, respectively.
These results are significantly lower than the experimental results: $T_{\rm h} \approx 85$ K \cite{Kubota,com} and 100 K \cite{Do}.

In each model, $R^z(T)$ increases for $T>T_{\rm h}$ and starts to saturate at $T\lessapprox T_{\rm h}$. 
As $R^z(T)$ increases with decreasing temperature, the intensity of the DSF $S(\mbox{\boldmath $0$},\omega;T)$ also starts increasing, as shown in Fig. \ref{fig2}.
Thus the growth of the $\Gamma$-point intensity of the DSF with decreasing temperature corresponds to the growth of the NN spin--spin correlation function.  
In contrast, the growth of the long-range spin correlation depends on the model used to calculate it.
The NN spin correlation and the long-range spin correlation increase separately for models 1, 3, and 4:
$S^{x}_{\boldsymbol Q}(T)$ at the magnetic ordering wave vector starts developing at $T \approx T_{\rm h}$, where the NN spin correlation is almost saturated.
$S^{x}_{\boldsymbol Q}(T)$ in models 1 and 3 saturates at $T_{\rm \ell}$, while $S^{x}_{\boldsymbol Q}(T)$  in model 4 saturates at the temperature where the shoulder appears in $C(T)$.
Therefore, the low-temperature peak in models 1 and 3 and the shoulder in model 4 are interpreted to be indicative of a release of entropy as the system moves toward the zigzag-ordered phase.
In model 2, $S^x_{\boldsymbol Q}(T)$ at the $Y$ point and $R^z(T)$ increase simultaneously as the temperature decreases and both saturate slightly below $T_{\rm h}$. 
Consequently, the origin of the low-temperature peak in model 2 seems to differ from that of the other models.

\subsection{Discussion of numerical results for models 1--4}

Concerning the constant-energy cuts shown in Fig. \ref{fig4}, models 1 and 4 explain the INS experiments qualitatively, whereas models 2 and 3 do not explain the INS experiments. 
The former may be reasonable for model 4 because the interactions in model 4 are tuned so as to reproduce the low-energy excitations of $\alpha$-RuCl$_3$~\cite{Winter2}. 
Models 1--3 reproduce the two-peak structure of $C(T)$ observed in the experiments.
Of the effective models that explain the magnetic zigzag ground state, 
models 1 and 4 clearly lead to separate growth of the NN spin correlation and of the long-range spin correlation.

We consider that the separate growth of both the NN spin correlation and the long-range spin correlation is significant for explaining the INS and $C(T)$ experiments.
The intensity at the $\Gamma$ point observed in the INS experiments above 100 K~\cite{Do,Banerjee2} is evidence of the separate growth of the NN spin correlation and the long-range spin correlation.
In fact, the magnetic Bragg peak starts to grow below $T_N$~\cite{Banerjee1,SYPark}.
In this section, we have demonstrated that the growth of the $\Gamma$-point intensity of the DSF is associated with the growth of the NN spin correlation at $T \approx T_{\rm h}$.
This value is comparable to the energy scale of the dominant Kitaev interaction.

The problem with models 1 and 4 is that $T_{\rm h}$ is too low: the values of $T_{\rm h}$ for models 1 and 4 are about one-fourth and one-seventh of the experimentally observed value $T_{\rm h} \approx 85$ K  \cite{Kubota,com}, which is the most optimistic case.
A characteristic temperature is also present in experiment at approximately 100 K~\cite{DHirobe}, and the thermal conductivity also exhibits an additional peak at approximately 100 K~\cite{DHirobe}. 
The authors in Ref. \onlinecite{DHirobe} have claimed that this additional conduction is rather immune to the structural phase transition, which suggests that it does not originate from phonons.
Thus the experimentally observed high-temperature peak in $C(T)$ is an intrinsic property of the magnetic part of $\alpha$-RuCl$_3$ and is explained within the effective magnetic model.
This result is due to the energy scale of the spin--orbit interaction and of the Hund coupling being of the order ${\mathcal O}(10^3)$ K.
From our numerical calculations performed so far for many sets of interactions, 
we find that $T_{\rm h}$ of $C(T)$ is mainly controlled by the largest Kitaev interaction.  
These findings suggest that the NN ferromagnetic Kitaev interaction is underestimated in models 1--4.

\section{\label{sec:level4} Empirical model with strong Kitaev interaction}
\subsection{Possible empirical model for $\alpha$-${\rm RuCl_3}$}
\begin{figure*}[htb]
\begin{center}
\includegraphics[width=\hsize]{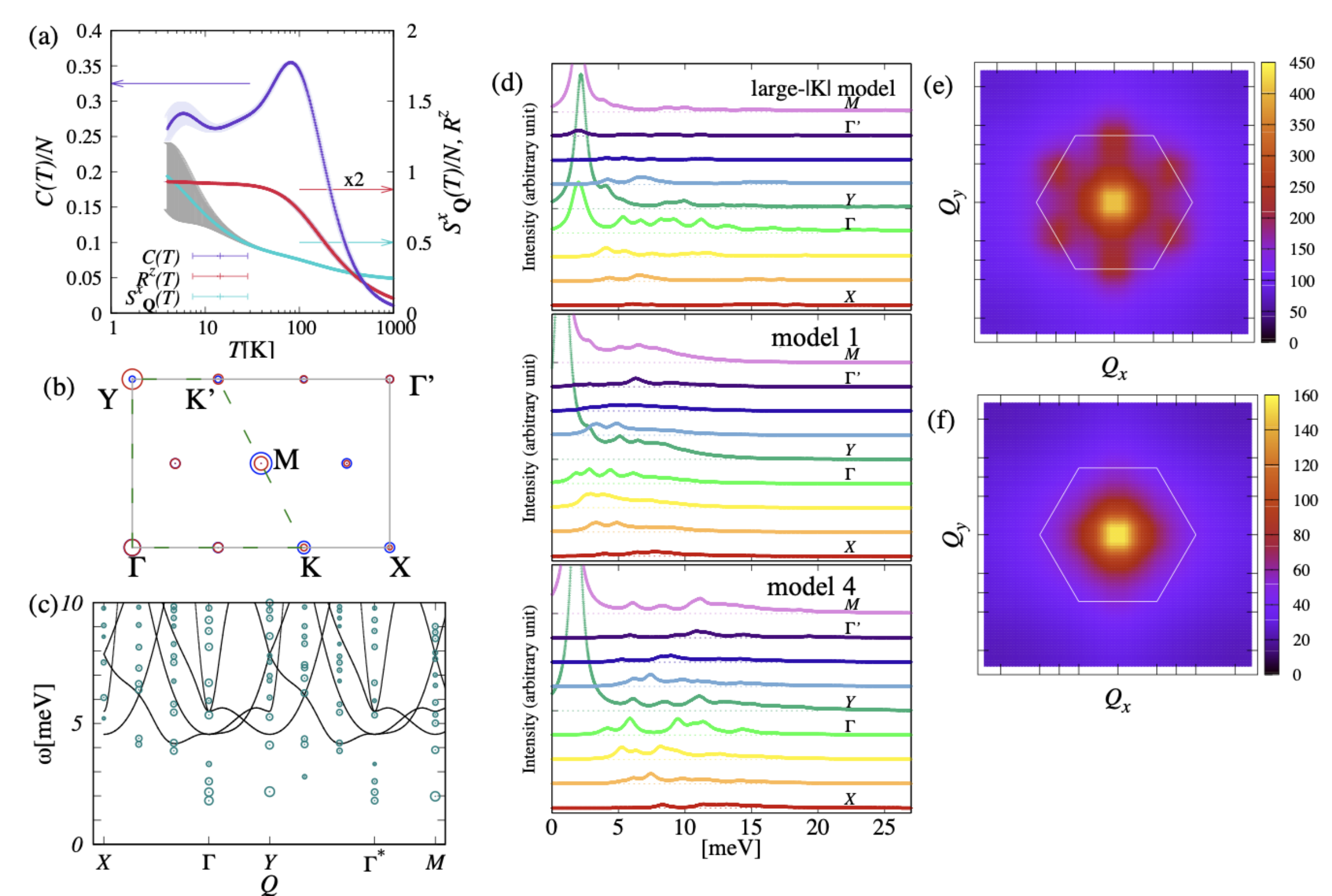}
\vspace{0pc}
\hspace{0pc}
\caption{\label{fig_strongK} Results of the large-$|K|$ model. (a) Temperature dependences of heat capacity $C(T)$, the longitudinal component $R^z(T)$ of the NN spin--spin correlation function, 
and the static spin structure factor $S^x_{\boldsymbol Q}(T)$ at the $Y$ point calculated by using thermal pure quantum states.
(b) Static spin structure factor $S^{x/z}_{\boldsymbol Q}(T=0)$ in a zero magnetic field.
(c) $S({\boldsymbol Q},\omega;T=0)$ in a zero magnetic field. 
The area of the circle represents $\log S({\boldsymbol Q},\omega;T=0)$ and the half width of the Lorentzian is set to 0.001 meV. 
The solid curves are the result obtained by LSWT. 
(d) $S({\boldsymbol Q},\omega;T=0)$ in a zero magnetic field. 
The results are drawn with the sum of the Lorentzian peaks with the half width 0.5 meV.
The curves from the bottom to the top correspond to the results at the wave numbers traveling from the $X$ point to the $M$ point along the arrows shown in Fig. 1 (d).
Top panel is the result for the large-$|K|$ model.  The results for models 1 and 4 are shown in the middle and bottom panels, respectively.
(e) and (f) Constant-energy cuts integrated over energy windows [1.5,3] meV and [9,12] meV, respectively.  
The half width of the Lorentzian in $S({\boldsymbol Q}, \omega;T=0)$ is set to 0.25 meV, which is comparable to the resolution of the INS experiments~\cite{Do,Banerjee2}. 
All results are calculated by using $N=24$ clusters.}
\end{center}
\end{figure*}

In Sec. \ref{sec:level3}, we have argued that the NN ferromagnetic Kitaev interactions are underestimated in models 1--4. 
Although all four models are in the zigzag-ordered phase,
only models 1 and 4 explain the constant-energy cuts of the scattering intensity.
Thus we consider that the combination of the NN interactions, namely, the ferromagnetic Kitaev interaction, the positive $\Gamma$, and the weak ferromagnetic Heisenberg interaction, is important to reproduce the INS experiments.
Between models 1 and 4, we focus on model 1 because it reproduces the two-peak structure in $C(T)$.
In model 1, we change the NN Kitaev interactions into strong interactions, while maintaining the other interactions unchanged. 
We set $K_z=-40$ meV and $K_{x/y}=-38.57$ meV, retaining the same ratio $K_z/K_{x/y} \approx 1.037$ as in model 1. 
The value of $K_z$ is set such that the calculated $T_{\rm h}$ agrees with the experimental values~\cite{Kubota,com,Do}. 
We call this model the large-$|K|$ model.

Figure \ref{fig_strongK}(a) shows the temperature dependence of $C(T)$, $R^z(T)$, and $S^x_{\boldsymbol Q}(T)$ at the $Y$ point for the large-$|K|$ model.
The large-$|K|$ model reproduces the two-peak structure in $C(T)$, and
the peak temperatures are $T_{\rm h} \approx$ 80 K  and $T_{\rm \ell} \approx$ 5 K, which are
both quite close to the experimentally observed values~\cite{Kubota,com,Do}.  
The important point is that the energy scale of $T_{\ell}$ hardly changes, although we increase the Kitaev interaction by a factor of approximately six. 
This is reasonable because the magnetic ordering expected approximately below $T_{\ell}$ is mainly stabilized by the $\Gamma$ term and the weak NN Heisenberg interaction.
Conversely, $T_{\rm h}$ shifts to the high-temperature region thanks to the more negative Kitaev interaction.
We observe the separate growth of the NN spin--spin correlation function $R^z(T)$ and the long-range spin correlation $S^x_{\boldsymbol Q}(T)$ characterizing the magnetic zigzag order:
$R^z(T)$ saturates at $T \approx T_{\rm h}$ and the phase transition to the magnetic zigzag order is expected at $T \lessapprox T_{\rm \ell}$, where $S^x_{\boldsymbol Q}(T)$ at the $Y$ point starts to saturate.
These results show that the large-$|K|$ model is consistent with the $C(T)$ experiments not only qualitatively but also quantitatively.

Figures \ref{fig_strongK}(b) and \ref{fig_strongK}(c) show the SSF and DSF at $T=0$ for the large-$|K|$ model.
The ground state of the large-$|K|$ model remains in the zigzag-ordered phase because the largest intensity of the SSF appears at the $Y$ and $M$ points. 
However, the system is expected to be near the KSL phase because of the strong Kitaev interaction.
The LSWT curves shown in Fig. \ref{fig_strongK}(c) are located in the higher-energy region and thus do not explain the low-lying excitation of the DSF.
In Fig. \ref{fig_strongK}(d), we show the intensity profiles of the DSFs on a linear scale.
Except for the low-lying excitations at the $\Gamma$, $Y$, and $M$ points, the broad peaks appear in 4 meV $\lessapprox \omega \lessapprox$ 20 meV in the large-$|K|$ model. 
In models 1 and 4,  the broad peaks are observed in 3 meV $\lessapprox \omega \lessapprox$ 15 meV and 4 meV $\lessapprox \omega \lessapprox$ 20 meV, respectively.
The presence of the broad peaks means that, in these three models, the excited states are distributed densely in these energy regions and compose the excitation continuum in the thermodynamic limit.
Even though the absolute value of the Kitaev interaction increases approximately six times in the large-$|K|$ model,
we consider that the density distribution of the excited states is similar to that in models 1 and 4.

Figures \ref{fig_strongK}(e) and \ref{fig_strongK}(f) show the constant-energy cuts of the scattering intensity ${\mathcal I}({\boldsymbol Q},\omega; T=0)$ at $T=0$ for the $N=24$ clusters. 
The half width of the Lorentzian is set to 0.25 meV, which is comparable to the resolution of the INS experiments~\cite{Do,Banerjee2}. 
The constant-energy cut integrated over the energy window [1.5,3] meV shows the remarkable six-pointed star-shaped intensity with a large intensity at the $\Gamma$ point, whereas the same integrated over the energy window [9,12] meV forms concentric circles of intensity centered at the $\Gamma$ point. 
These results are consistent with the typical features of the INS experiments~\cite{Do,Banerjee2}.
Thus we argue that the large-$|K|$ model is more appropriate for explaining the INS and $C(T)$ experiments than are models 1--4.

\subsection{Lowest-energy states of large-$|K|$ model in magnetic fields}
\begin{figure}[htb]
\begin{center}
\includegraphics[width=0.75\hsize]{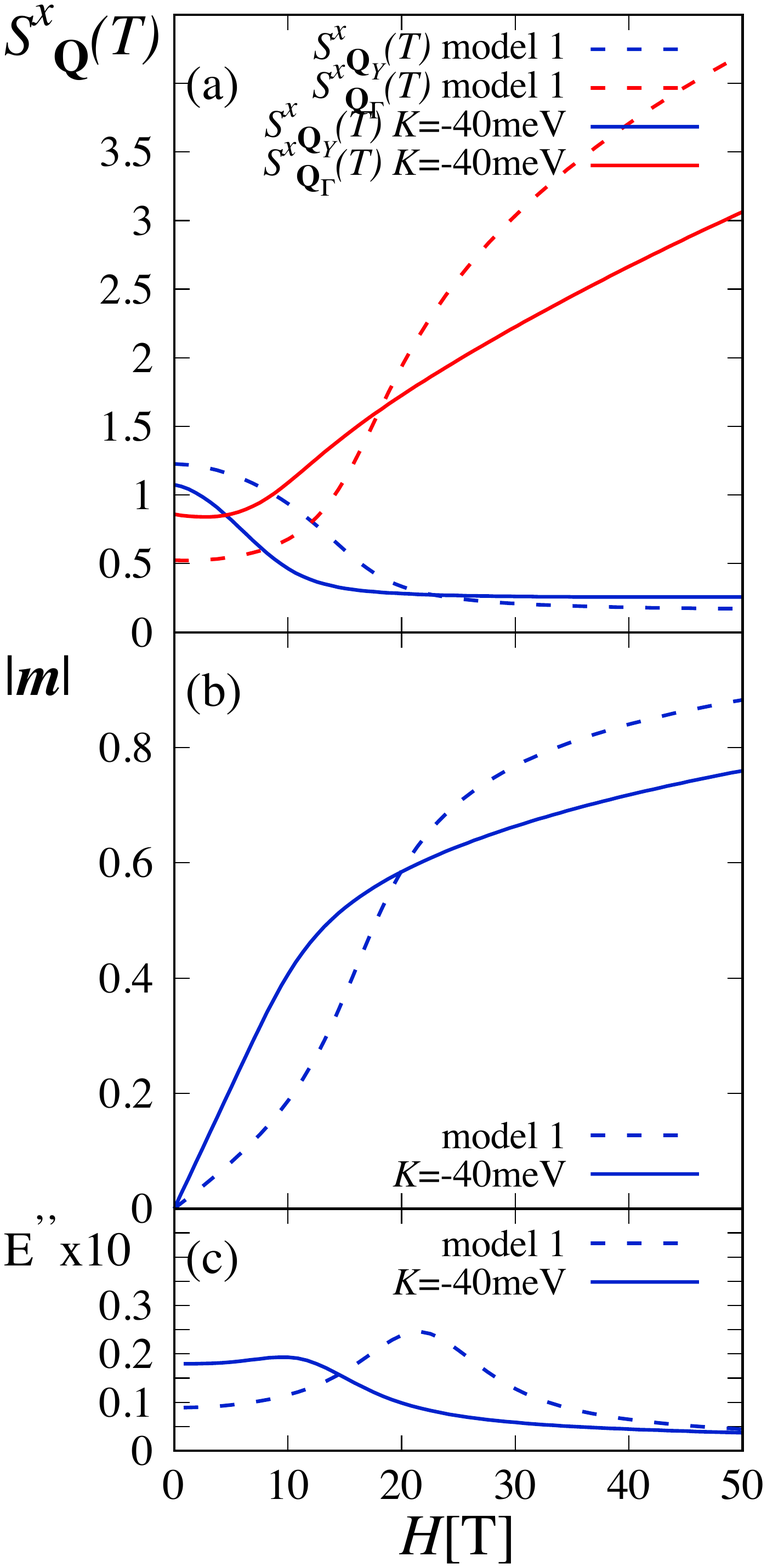}
\vspace{0pc}
\hspace{0pc}
\caption{\label{fig_field_dep} (a) Field dependence of static spin structure factors $S^x_{\boldsymbol Q}(T=0)$ and (b) magnetization curves at zero temperature.  
In-plane magnetic fields are applied to the $N=24$ cluster with the field direction parallel to the $b$ axis shown in Fig. 1(a).
Solid (dotted) curves show the results for the large-$|K|$ model (model 1). 
The field dependence of $S^x_{\boldsymbol Q}(T=0)$ at the $\Gamma$ ($Y$) point is shown by the red (blue) curve. 
Following the discussion in Refs.~\onlinecite{Winter} and \onlinecite{Yadav}, we set the in-plane $g$ value to 2.3.
 The saturated magnetization in panel (b) is normalized to unity. 
(c) Numerical second derivative of internal energy, ${\rm E}''=\frac{\partial^2 E}{{\partial H}^2}$.}
\end{center}
\end{figure}

We now use the large-$|K|$ model to investigate $\alpha$-${\rm RuCl_3}$ with applied magnetic fields. 
The Hamiltonian is 
\begin{eqnarray}
{\mathcal H} = {\mathcal H}_{\rm KH} 
-\mu_{\rm B} \sum_{i} \sum_{\mu, \nu=x,y,z} H^{\mu}  \hat{g}^{\mu \nu} S^{\nu}_i,
\label{Ham3}
\end{eqnarray}
where ${\mathcal H}_{\rm KH}$ is the generalized form of the Kitaev--Heisenberg Hamiltonian given in Eq. (\ref{Ham2}), $\mu_{\rm B}$ is the Bohr magneton, and $\hat{g}^{\mu \nu}$ denotes the $g$ tensor. 
Following the discussion in Refs.~\onlinecite{Winter} and \onlinecite{Yadav}, we set the in-plane $g$ value to 2.3.
We assume that the magnetic fields are oriented parallel to the ${\boldsymbol b}$ axis in Fig. \ref{fig5}(d), which corresponds to the experimental conditions~\cite{Do,Banerjee3,ZWang}.

Figure \ref{fig_field_dep}(a) shows the field dependence of the SSFs and the magnetization curves at $T=0$.
Upon increasing the in-plane magnetic field $H=|{\boldsymbol H}|$ (${\boldsymbol H}  || {\boldsymbol b}$), the wave vector of the dominant intensity in $S^x_{\boldsymbol Q}(T=0)$ switches from the $Y$ point to the $\Gamma$ point at $H_{\gamma} \approx 5$ T. 
This result indicates that the magnetic zigzag order is suppressed for $H \gtrapprox H_{\gamma}$. 
Note that this value is almost independent of the system size. 
The suppression field (i.e., the field strength required to suppress the magnetic order) is less than that obtained with model 1. 
In model 1, the wave vector of the dominant intensity of $S^x_{\boldsymbol Q}(T=0)$ changes from the $Y$ point to the $\Gamma$ point at $H \approx 13$ T, which means that the magnetic zigzag order is suppressed for $H \gtrapprox 13$ T. 
Experiments show that the magnetic zigzag order is suppressed at $H \approx 7$ T~\cite{SHBaek,ANPonomaryov,Banerjee3,JASears,IALeahy,AUBWolter} for an in-plane magnetic field.
Although the large-$|K|$ model predicts a suppression field that is slightly less than the experimentally obtained field, the agreement is quantitatively satisfactory.

Figure~\ref{fig_field_dep}(b) shows the magnetization curves for the large-$|K|$ model and for model 1. 
The magnetization curve for the large-$|K|$ model increases almost linearly up to $H_s \approx 11$ T and then gradually saturates, 
whereas the magnetization curve for model 1 makes a clear ``S-shaped'' curve. 
As the Kitaev interaction is made increasingly negative with respect to model 1, the ``S'' shape of the magnetization curve for model 1 changes continuously to become almost linear, which means that the magnetization curves of the large-$|K|$ model and of model 1 are the topologically equivalent,
although the magnetization curve for the large-$|K|$ model appears almost linear at this resolution.
In experiments\cite{Sears,Kubota,RDJohnson,Majumder}, the magnetization rapidly increases up to $H \approx 13$ T, making a very weak S-shaped curve and revealing a weak anomaly near $H \approx 8$ T~\cite{Sears,RDJohnson}.
Above $H \approx 13$ T, the magnetization curve increases gradually toward the saturation value.
Although it is difficult to discuss the anomaly in the magnetization curves for the large $|K|$ model, 
the topology of the magnetization curves is equivalent with each other and thus, the large $|K|$ model qualitatively explains the experimental results.
The gradual increase of the magnetization in the high field is due to the off-diagonal terms in the Hamiltonian. 
This situation is similar to that in systems with the Dzyaloshinskii--Moriya interactions\cite{DM1,DM2}, which is expressed as antisymmetric off-diagonal terms. 
Thus we expect the classical fully polarized state to appear for $H \gtrapprox H_s$ (${\approx} 11$ T).

To see the phase-transition properties, Fig.~\ref{fig_field_dep}(c) shows the second derivative ${\rm E''}= \frac{\partial^2 E}{\partial H^2}$ of the internal energy $E$ with respect to the magnetic field for the large-$|K|$ model. 
Upon applying an in-plane magnetic field, the current system size makes it difficult to find a clear peak separating the zigzag-ordered phase from the higher-field phase.
However, we consider that the magnetic zigzag order is strongly suppressed above $H_{\gamma} \approx 5$ T because, at these field strengths, the dominant intensity of the SSF shifts from the $Y$ point to the $\Gamma$ point.
In contrast, a peak appears around $H_s$, but whether this peak is related to the second-order phase transition is difficult to determine because of the results for small-sized clusters.
We therefore consider that the fully polarized state is stable for $H \gtrapprox H_s$.
In particular, the first derivative of the magnetization curves rapidly decreases at this field strength and the magnetization continuously increases toward saturation with no singularity.

\subsection{Polarized THz spectra of large-$|K|$ model in magnetic fields}

\begin{figure}[htb]
\vspace{0pc}
\hspace{2pc}
\begin{center}
\includegraphics[width=0.75\hsize]{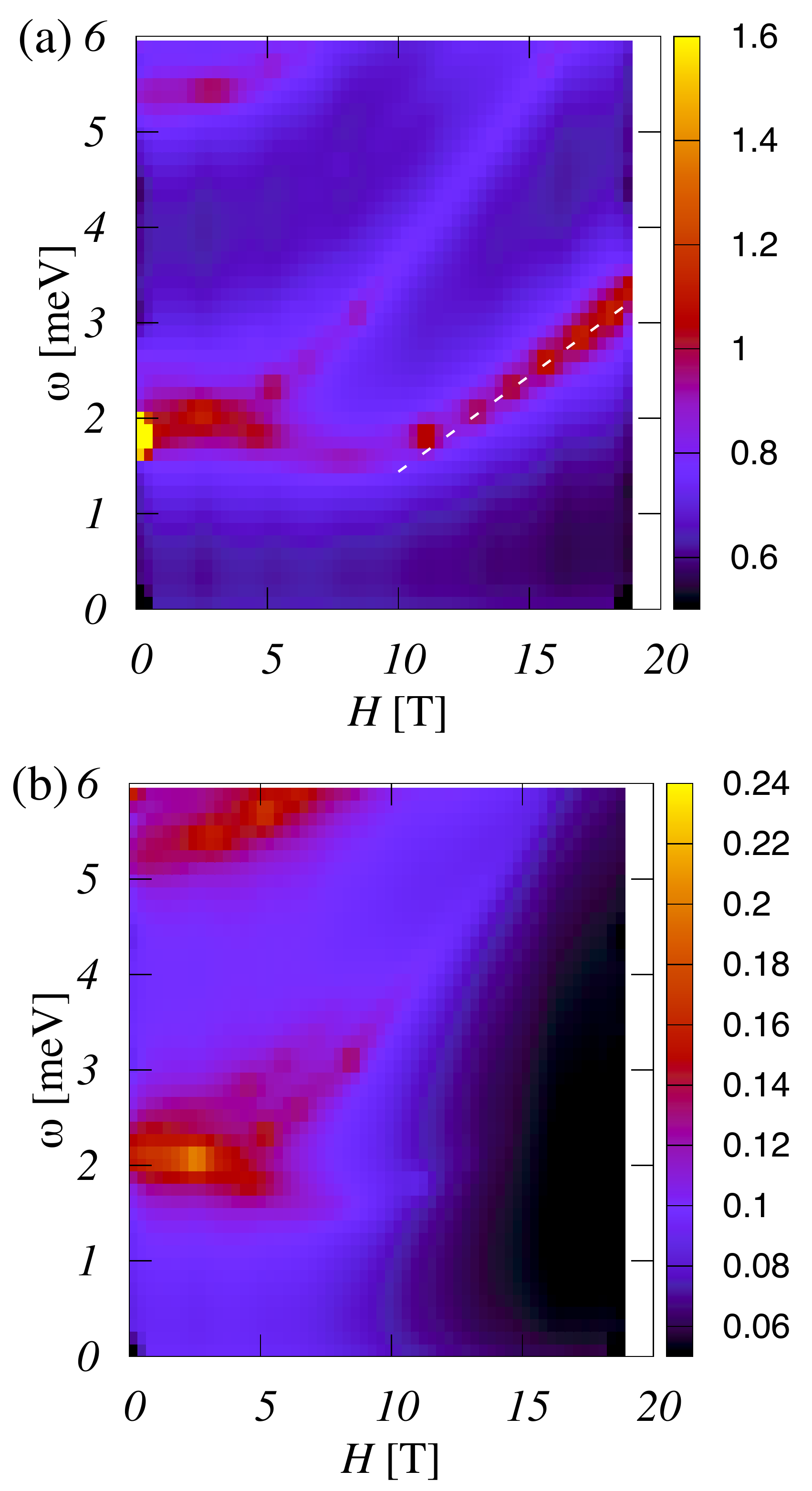}
\caption{\label{ESR} Polarized THz spectra, $\omega \chi''(\omega) \propto \omega S(\Gamma,\omega)$~\cite{Little}, with ${\boldsymbol H}||{\boldsymbol b}$. (a) ${\boldsymbol h}^{\omega} \parallel {\boldsymbol H}$  and (b)  ${\boldsymbol h}^{\omega} \perp {\boldsymbol H}$. All results are calculated by using the $N=24$ cluster with the Lorentzian half width set to 0.01 meV to highlight the peak positions. The dotted line shows the slope with 0.22 meV/T. 
Note that the upper limit of the intensity differs in panels (a) and (b).}
\end{center}

\end{figure}

Figure \ref{ESR} shows the field dependence of the polarized THz spectra, $\omega \chi''(\omega) \propto \omega S(\Gamma,\omega)$~\cite{Little}, which are measured in the form of absorption coefficients in THz  spectroscopy.
The signal for ${\boldsymbol h}^{\omega} \perp {\boldsymbol H}$ is stronger than that for ${\boldsymbol h}^{\omega} \parallel {\boldsymbol H}$, 
where ${\boldsymbol h}^{\omega}$ is the magnetic field of the electromagnetic THz  wave. 
At zero field, a spin-gap excitation appears at $\omega \approx 2$ meV in both components ${\boldsymbol h}^{\omega} \perp {\boldsymbol H}$ and ${\boldsymbol h}^{\omega} \parallel {\boldsymbol H}$. 
As the magnetic field increases up to $H_s \approx 11$ T, these lowest-energy signals form convex curves without closing the energy gap.
For $H \gtrapprox H_s$, the lowest-energy signal becomes linear as a function of $H$ for ${\boldsymbol h}^{\omega} \parallel {\boldsymbol H}$.
The excitation energy increases at the rate ${\approx} 0.22$ meV/T.
For ${\boldsymbol h}^{\omega} \perp {\boldsymbol H}$, the signals of the lowest-energy mode are strongly suppressed for $H \gtrapprox H_s$ and a line appears with a larger slope.
We considered that the $\Delta S =1$ mode appears for ${\boldsymbol h}^{\omega} \parallel {\boldsymbol H}$ and the prominent mode for ${\boldsymbol h}^{\omega} \perp {\boldsymbol H}$ accompanies the larger-$\Delta S$ excitation.
These results are qualitatively consistent with recent THz spectroscopy measurements~\cite{ZWang}, which indicate that the spin gap $\omega \approx 2.5$ meV, and the slope $\approx 0.38$ meV/T\cite{Comment}.

The linear response of the lowest mode for ${\boldsymbol h}^{\omega} \parallel {\boldsymbol H}$ with $H>H_s$ is also obtained with model 4~\cite{Winter2}.
It has been argued in Ref.~\onlinecite{Winter2}  that no regime exists where $Z_2$ fluxes are diluted, which hampers possible connections to the Kitaev's exact solution.
In the present model, the Kitaev coupling is quite large.
Nonetheless, we still observe a linear response in the polarized THz spectra, which seems to result from the classical fully polarized state.
The energy scale of the excitation modes in the above THz spectra is significantly less than that of the Kitaev interactions.
The result obtained thus implies that the low-energy magnetic excitations are governed not by the amplitude of the Kitaev coupling, but by the off-diagonal terms and the weak NN Heisenberg coupling.
Estimating the Kitaev interaction in $\alpha$-RuCl$_3$ would require investigations into thermal properties, such as the temperature dependence of $T_1^{-1}$ in NMR measurements, thermal transport, and $C(T)$.


\section{\label{sec:level5} Discussions and Summary}

We have calculated the DSFs and temperature dependencies of the heat capacity, the NN spin--spin correlation function, and the SSF for the three {\it ab initio} models and the {\it ab initio}-guided model.  
The results of the calculations have shown that the INS feature at zero field is qualitatively described by models 1 and 4.
The two-peak structure of the heat capacity is qualitatively reproduced by models 1--3, whereas the low-temperature peak is absent and a shoulder appears when using model 4.   
Thus models 1 and 4 qualitatively describe the experimental features of $\alpha$-RuCl$_3$.
The disadvantage of models 1 and 4 is that they place $T_{\rm h}$ significantly below the temperature obtained experimentally~\cite{Do,Kubota}.
Thus we have concluded that the NN Kitaev coupling is underestimated in the {\it ab initio} calculations.

To explain the peaks in heat capacity as a function of temperatures in addition to the INS experiments,
we have proposed the large-$|K|$ model with parameters equivalent to those used in model 1, except for the NN Kitaev coupling.
The results for the large-$|K|$ model are qualitatively and quantitatively consistent with both experimental features.
We have demonstrated that the large-$|K|$ model also reproduces experimental results of polarized THz spectra in a magnetic field.
These results have suggested that the magnetic excitation in the low-energy region is essentially characterized not by a dominant Kitaev interaction but by the $\Gamma$ term and the weak Heisenberg interaction between NN pairs.

The large-$|K|$ model has a potential to explain the Raman experiments\cite{LJSandilands}.
As shown in Fig. 7(d), the broad peaks in the DSF appear approximately in 4 meV $\lessapprox \omega \lessapprox$ 20 meV in the large-$|K|$ model.
These broad peaks originate from the excitation continuum in the thermodynamic limit, which means that the excited states are distributed densely in 4 meV $\lessapprox \omega \lessapprox$ 20 meV.
We consider that these excited states contribute to the broad Raman spectra.
In the Raman experiments\cite{LJSandilands}, it has been shown that the excitation continuum extends up to 20 $\sim$ 25 meV.
Thus the large-$|K|$ model is not inconsistent with the Raman experiments~\cite{LJSandilands}.

In the large-$|K|$ model, $K_z=-40$ meV seems to be large when we consider the realistic model.
We have also performed the same computations for $K_z=-30$ meV.
However, not only the DSF at zero field but also the THz spectra in the magnetic fields still reproduce experimental results qualitatively, for which $T_{\rm h}$ shifts to the lower value as expected.


Note that it remains unclear why the current {\it ab initio} calculations provide the weak Kitaev coupling. 
{\it Ab initio} calculations involve empirical or tuning parameters such as the Hund coupling and, in some cases, Coulomb interactions.
Thus we may obtain a more-negative Kitaev coupling $K$ that would be comparable to that of ${\rm Na_2IrO_3}$~\cite{Yamaji}. 
The importance of the combination of interactions, 
namely the ferromagnetic Kitaev interactions and the positive $\Gamma$ terms for the nearest neighbor pair, has been emphasized in recent works~\cite{Yadav,Winter,KRan,WWang,Kim,Winter2,Winter3,LJanssen,ACatuneanu}.
We believe that further {\it ab initio} calculations and discussions are highly desired to obtain a proper model of $\alpha$-RuCl$_3$.
In addition, further experiments and analyses of the thermal properties, such as the temperature dependence of $T_1^{-1}$ in NMR measurements, thermal transport, and so on, are also needed to obtain the essential information required to estimate the Kitaev coupling.

\begin{acknowledgments}
We thank M. Imada, N. Kawashima, H. Tanaka, and Y. Yamaji for fruitful discussions.  
We also thank L. Janssen and S. Winter for valuable comments.
This work was supported by the Creation of new functional Devices and high-performance Materials to Support next-generation Industries (CDMSI), Challenge of Basic Science - Exploring Extremes through Multi-Scale Simulations (CBSM2), and KAKENHI (Grants No. 15K05232 and No. 16K17751) from MEXT, Japan. 
T.S. thanks the computational resources of the K computer provided by the RIKEN Advanced Institute for Computational Science through the HPCI System Research project (hp160201, hp160262, hp170262, and hp170263). 
We are also grateful for the numerical resources at the ISSP Supercomputer Center at the University of Tokyo  
and the Research Center for Nano-micro Structure Science and Engineering at University of Hyogo.

\end{acknowledgments}


\clearpage
\newpage
\pagebreak

\onecolumngrid
\begin{center}
  \textbf{\large Erratum: Effective model with strong Kitaev interaction for $\alpha$-${\rm RuCl_3}$ [Phys. Rev. B  {\bf 97}, 134424 (2018)]}\\[.2cm]
  Takafumi Suzuki$^1$ and Sei-ichiro Suga$^1$\\[.1cm]
  {\itshape ${}^1$Graduate School of Engineering, University of Hyogo, Himeji 671-2280, Japan\\}
  ${}^*$Electronic address: takafumi-s@eng.u-hyogo.ac.jp\\
(Dated: \today)\\[1cm]
\end{center}
\twocolumngrid

\setcounter{equation}{0}
\setcounter{figure}{0}
\setcounter{table}{0}
\setcounter{page}{1}
\renewcommand{\theequation}{E\arabic{equation}}
\renewcommand{\thefigure}{E\arabic{figure}}
\renewcommand{\bibnumfmt}[1]{[#1]}
\renewcommand{\citenumfont}[1]{#1}

\begin{figure}[htb]
\hspace{0pc}
\begin{center}
\includegraphics[width=0.5\hsize]{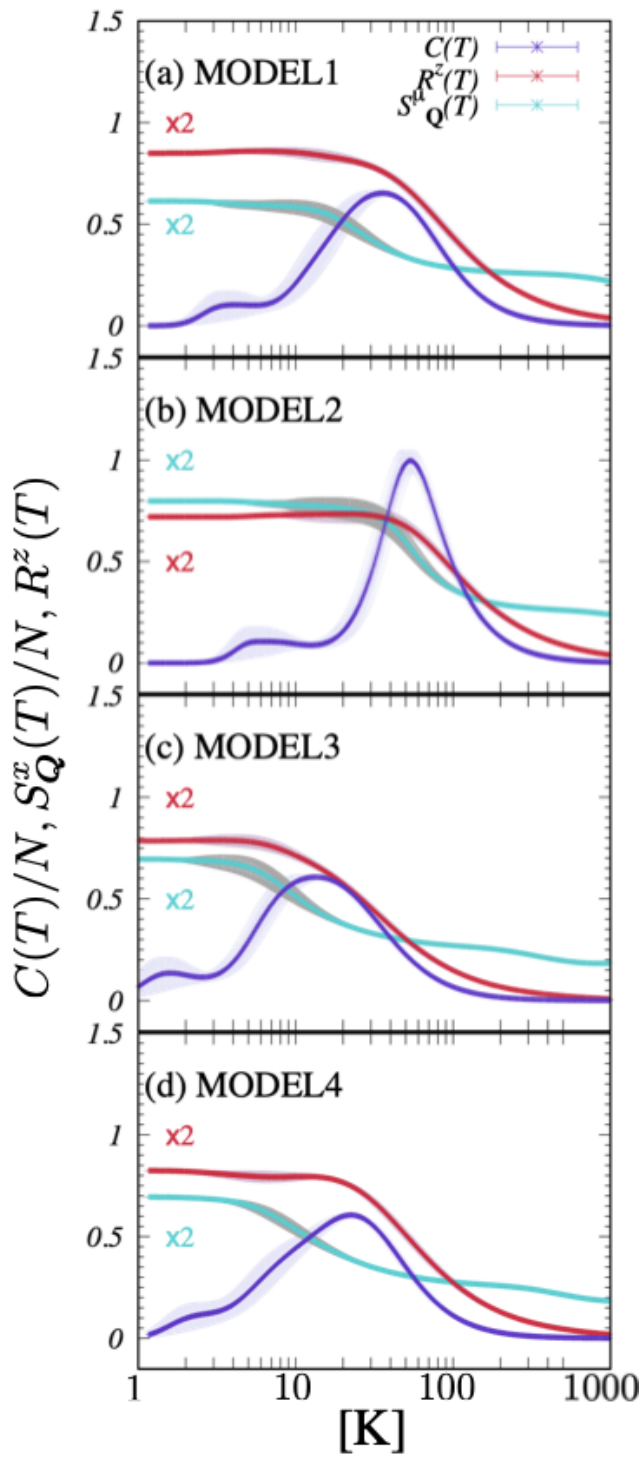}
\caption{\label{fig1} Corrected Fig. 6. 
Temperature dependence of heat capacity $C(T)$, longitudinal component $R^z(T)$ of the nearest-neighbor spin--spin correlation function, and static spin structure factor $S^{x}_{\boldsymbol Q}(T)$ at the $M$ point. 
}
\end{center}
\end{figure}

We have found the errors in the original article. Temperatures of the calculated quantities have to be multiplied by 2. We correct our previous results in order.

Figure \ref{fig1} shows the corrected $C(T)$, $S^{x}_{\boldsymbol Q}(T)$, and $R^z(T)$ for models 1--4. The high-temperature peak in $C(T)$ appears at  $T_{\rm h} \approx$ 40, 56, 14, and 24 K, respectively. 
These corrections do not affect the conclusion about models 1--4.

\begin{table}[thb]
 \caption{\label{table1} Coupling constants for the corrected large-$|K|$ model. 
Energy is expressed in meV.
}
\begin{tabular}{cccccccc}
\hline \hline 
               $J_{\rm 1st}^{x/y}$ & $J_{\rm 1st}^{z}$ & $K_{\rm 1st}^{x/y}$ & $K_{\rm 1st}^{z}$  
            & $\Gamma_{\rm 1st}^{x/y}$ & $\Gamma_{\rm 1st}^{z}$  
            & $\Gamma_{\rm 1st}^{\prime x/y}$ & $\Gamma_{\rm 1st}^{\prime z}$ \\ 
            \hline
            -1.55 & -1.49 & -24.11 & -25 & 5.24 & 5.28 & -1.08 & -0.69  \\
\hline\hline
\\
\end{tabular}
\end{table}
The coupling constants of the large-$|K|$ model are reset in the same way so as to reproduce $T_{\rm h}$ and $T_{\ell}$ in the experimental values. We thus reset $K_z=-25$ and $K_{x/y}=-24.11$ meV. The coupling constants are listed in Table \ref{table1}.

\begin{figure*}[htb]
\begin{center}
\includegraphics[width=0.8\hsize]{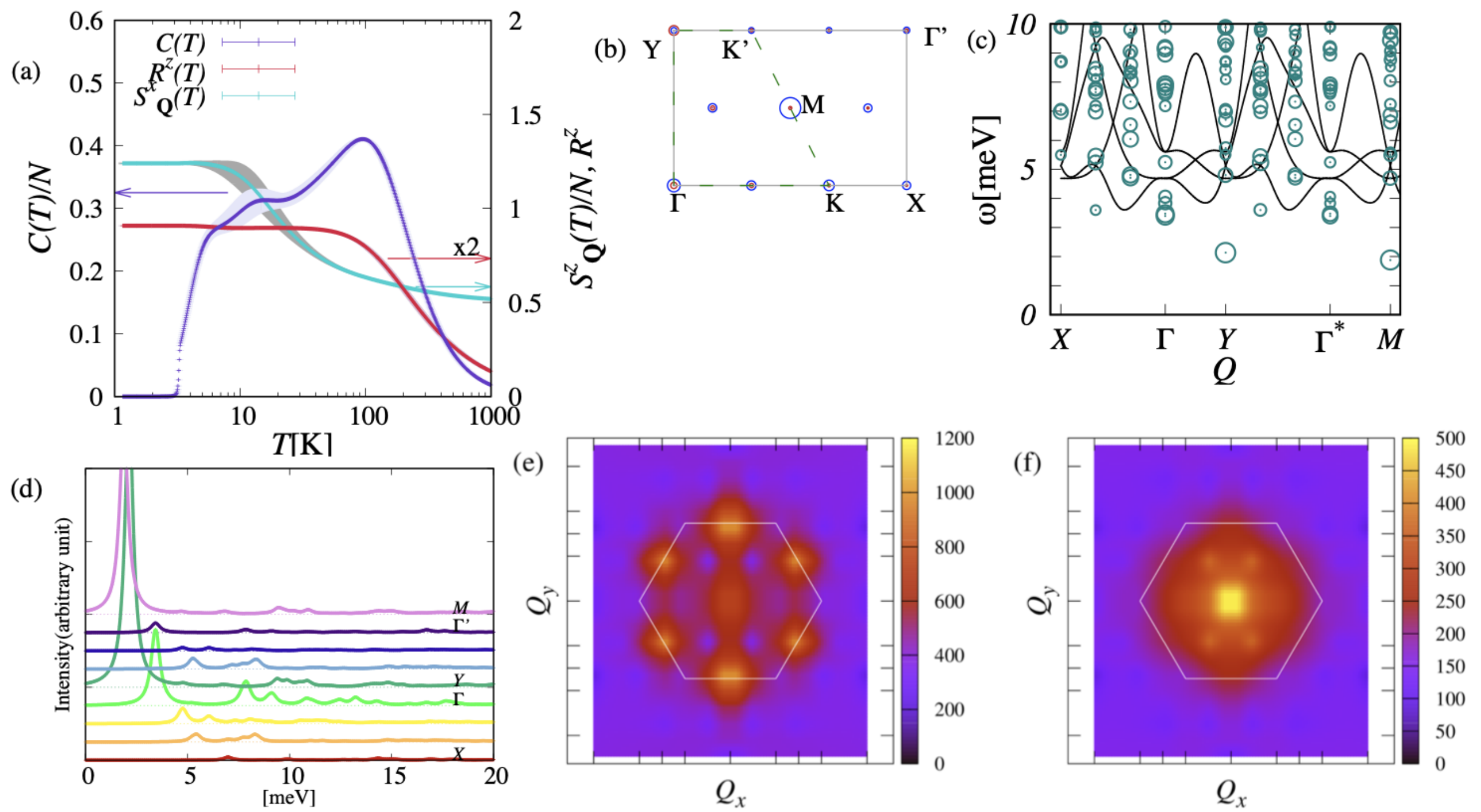}
\hspace{0pc}
\caption{\label{fig_strongK} Results for the corrected large-$|K|$ model. Corrected Fig. 7.
}
\end{center}
\end{figure*}
Figure \ref{fig_strongK}(a) shows the $C(T)$, $S^{x}_{\boldsymbol Q}(T)$, and $R^z(T)$ for the corrected large-$|K|$ model. $C(T)$ shows a two-peak structure at $T_{\rm h} \approx 100$ and $T_{\ell} \approx 12$ K, which are consistent with the experimental values~\cite{Kubota,Do}.
We also observe the separate growth of $R^z(T)$ and $S^z_{\boldsymbol Q}(T)$. At $T \approx T_{\rm h}$ $R^z(T)$ saturates and the phase transition to the magnetic zigzag order is expected at $T \lessapprox T_{\ell}$, where ${S^z}_{{\boldsymbol Q}}(T)$ at the $M$ point starts to saturate.

Figures \ref{fig_strongK}(b)-\ref{fig_strongK}(d) show the static spin structure factor (SSF) and dynamical spin structure factor (DSF) for the large-$|K|$ model.
The largest peak of the SSF appears at the $M$ point, which means that the ground state is in the zigzag phase.
Although the dispersion curves based on the linear spin-wave theory (LSWT) shift to the lower energy region, the LSWT curves do not explain the low-lying excitation in the DSF.

Figures \ref{fig_strongK}(e) and \ref{fig_strongK}(f) show the constant-energy cuts of the scattering intensity ${\mathcal I}({\boldsymbol Q},\omega; T=0)$ integrated over [1.5,3] meV and  [9,12] meV, respectively. 
The former shows the six-pointed star-shaped profile with a large intensity at the $\Gamma$ point. The latter shows the large intensity centered at the $\Gamma$ point. They reproduce the typical features of the inelastic neutron-scattering experiments \cite{Do,Banerjee2}.
These results sustain our conclusion that the low-energy excitation is governed mainly by the symmetric off-diagonal and Heisenberg interactions.

Figures \ref{ESR}(a)-\ref{ESR}(c) show the field dependence of $S^z_{\boldsymbol Q}(T=0)$, magnetization curve, and the second derivative of the internal energy, respectively. Upon increasing the in-plane magnetic field, the wave vector of the dominant intensity in $S^z_{\boldsymbol Q}(T=0)$ switches from the $M$ point to the $\Gamma$ point at $H \approx 8$ T. 
The magnetization curve makes a ``S-shape'' curve and increases gradually towards the saturation value above $H \approx 16$ T. 
The second derivative of the internal energy, $|E''|$, shows a peak at $H \approx 11$ T, meaning that the the fully polarized state appears above $H \approx 11$ T. 
If this magnetic field corresponds to the critical magnetic field where the zigzag magnetic order disappears,
the value is slightly higher than the experimental value. However, the results well explain the experiments~\cite{JASears,RDJohnson}. 

Figures \ref{ESR}(d) and \ref{ESR}(e) show the field dependence of the terahertz (THz) spectra. 
The field dependence of the THz spectra is the same as that for the previous large-$|K|$ model except for a spin-gap $\omega \approx 3.3$ meV at $H=0$ and the slope $\approx 0.24$ meV/T of a linear increase in the lowest-energy mode for $H\gtrapprox 16$ T in ${\boldsymbol h}^{\omega} \perp {\boldsymbol H}$. 

The corrected large-$|K|$ model does not affect the conclusions of the paper.

We thank Dr. S. Okamoto and Dr. P. Laurell for important suggestions.

\begin{figure}[htb]
\begin{center}
\includegraphics[width=0.8\hsize]{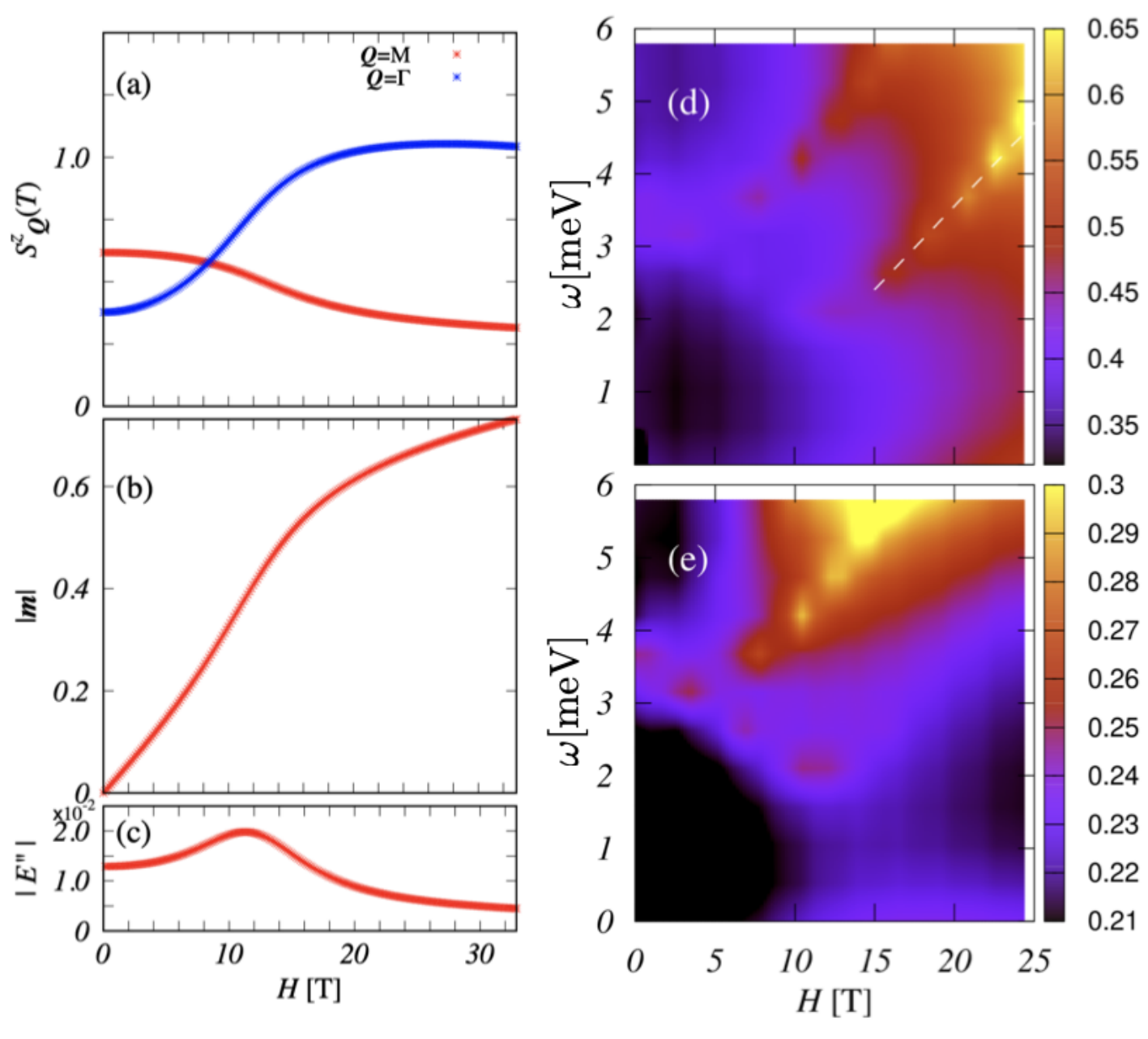}
\hspace{0pc}
\caption{\label{ESR}Results for the corrected large-$|K|$ model. (a)-(c) Corrected Fig. 8 and (d)-(e) corrected Fig. 9. 
}
\end{center}
\end{figure}


\begin{thebibliography}{99}
\bibitem{com} In Ref. \onlinecite{Kubota}, the two peaks in the temperature dependence of the heat capacity has not been discussed. However, we consider that the two peaks appear in Fig. 9 in Ref. \onlinecite{Kubota}. Using this result, we evaluate the higher peak to occur at a temperature of ${\approx}85$K.
\bibitem{Comment}
We have evaluated the slope from Fig. 3(d) in Ref. \onlinecite{ZWang}.
\end{thebibliography}
\end{document}